\documentclass[twocolumn]{emulateapj}  
\usepackage{times}

\usepackage{rotating}
\usepackage{graphicx,bm,amssymb}
\usepackage{epsfig}
\usepackage{amssymb}
\usepackage{natbib}
\usepackage{pstricks}
\usepackage{psfrag}
\usepackage[colorlinks=true,linkcolor=blue,citecolor=blue]{hyperref}
\usepackage{float}
\usepackage[caption = false]{subfig}
\def\apj{ApJ}
\def\apjl{ApJ}
\def\apjs{ApJS}
\def\aap{A\&A}
\def\mnras{MNRAS}
\def\prl{Phys.~Rev.~Lett.}
\def\nat{Nature}

\newcommand{\be}{\begin{equation}}
\newcommand{\ee}{\end{equation}}
\newcommand{\bary}{\begin{eqnarray}}
\newcommand{\eary}{\end{eqnarray}}

\shorttitle{GRB 190114C}
\shortauthors{Fraija N. et al.}
\begin{document}
\title{Synchrotron self-Compton as a likely mechanism of\\  photons beyond the synchrotron limit in GRB 190114C}
\author{N. Fraija$^{1\dagger}$, R. Barniol Duran\altaffilmark{2},  S. Dichiara\altaffilmark{3, 4} and P. Beniamini\altaffilmark{5} }
\affil{$^1$Instituto de Astronom\'ia, Universidad Nacional Aut\'{o}noma de M\'{e}xico, Apdo. Postal 70-264, Cd. Universitaria, M\'{e}xico DF 04510\\
$^2$ Department of Physics and Astronomy, California State University, Sacramento, 6000 J Street, Sacramento, CA 95819-6041, USA\\
$^3$ Department of Astronomy, University of Maryland, College Park, MD 20742-4111, USA\\
$^4$ Astrophysics Science Division, NASA Goddard Space Flight Center, 8800 Greenbelt Rd, Greenbelt, MD 20771, USA\\
$^5$ Department of Physics, The George Washington University, Washington, DC 20052, USA
}
\email{$\dagger$nifraija@astro.unam.mx}
%
%
\begin{abstract}
GRB 190114C, a long and luminous burst, was detected by several satellites and ground-based telescopes from radio wavelengths to GeV gamma-rays.  In the GeV gamma-rays, the Fermi LAT detected 48  photons above 1 GeV  during the first hundred seconds after the trigger time, and the MAGIC telescopes observed for more than one thousand seconds very-high-energy (VHE) emission above 300 GeV.  Previous analysis of the multi-wavelength observations showed that although these are consistent with the synchrotron forward-shock model that evolves from a stratified stellar-wind to homogeneous ISM-like medium,   photons above few GeVs can hardly be interpreted in the synchrotron framework.  In the context of the synchrotron forward-shock model,   we derive the  light curves and spectra of the synchrotron self-Compton (SSC) model in the stratified and homogeneous medium.   In  particular, we study the evolution of these light curves during the stratified-to-homogeneous afterglow transition.   Using the best-fit parameters reported  for GRB 190114C we interpret the  photons beyond the synchrotron limit in the SSC framework and model its spectral energy distribution.  We conclude that low-redshift GRBs described under a favourable set of parameters as found in the early afterglow of GRB 190114C could be detected at hundreds of GeVs, and also afterglow transitions would allow that VHE emission  could be observed for longer periods.
\end{abstract}

\section{Introduction}
Gamma-ray bursts (GRBs) are the most luminous explosions in the Universe, and one of the most promising sources for multimessenger observation of non-electromagnetic signals such as  very-high-energy (VHE) neutrinos, cosmic rays and gravitational waves.  Observation of sub-TeV photons from bursts would provide crucial information of GRB physics including hadronic and/or leptonic contributions, values of  the bulk Lorentz factors as well as  microphysical parameters.   In the Fermi Large Area Telescope (LAT) era, detections of bursts linked to GeV photons have been pivotal in painting a comprehensive picture of GRBs.\\
Recently, \cite{Ajello_2019} reported the second Fermi-LAT catalog which summarized the temporal and spectral properties of the 169 GRBs with high-energy photons above 100 MeV detected from 2008 to 2018.  Among the highest energy photons associated (with high probability $> 0.9$) with these bursts are: a  31.31-GeV photon arriving at 0.83 s after the trigger, a 33.39 GeV-photon at 81.75 s and a 19.56-GeV photon at 24.83 s, which were located around GRB 090510,  GRB 090902B and  GRB 090926A, respectively.  Beside this list,  GRB 130427A presented the highest energy photons ever detected, 73 GeV and 95 GeV  observed at 19 s and 244 s, respectively \citep{2014Sci...343...42A}, and GRB 160509A was related to a 52-GeV photon at 77 s after the trigger \citep{2016GCN.19413....1L}.  These bursts exhibited two crucial similarities: i) The first high-energy photon ($\geq 100$ MeV) was delayed with the onset of the prompt phase that was usually reported in the range of hundreds of keV and ii) The high-energy emission was temporarily extended, with a duration much longer than the prompt emission which was typically less than $\sim$ 30 seconds.\\
In the range of GeV and harder, VHE emission  is expected from the nearest and the brightest bursts. Alternative mechanisms to synchrotron radiation have been widely explored at internal as well as external shocks to interpret this emission.   Using hadronic models,  photo-hadronic interactions \citep{2009ApJ...705L.191A, 2000ApJ...537..255D, 2014MNRAS.437.2187F} and  inelastic proton-neutron collisions  \citep{2000ApJ...541L...5M} have been proposed.   However, the non-temporal coincidence between GRBs and neutrinos reported by the IceCube collaboration  have suggested that  the amount of hadrons are low enough so that hadronic interactions are non efficient processes \citep{2012Natur.484..351A, 2016ApJ...824..115A, 2015ApJ...805L...5A}.   Using  leptonic models,  external inverse Compton (EIC; \citealp{1996ApJ...471L..91P, 2000ApJ...544L..17P, 2018ApJ...859...70F}) and synchrotron self-Compton (SSC; \citealp{2001ApJ...559..110Z, 2012ApJ...755...12V,2001ApJ...546L..33W,2012ApJ...751...33F, 2012ApJ...755..127S,  2019ApJ...871..123F,  2019arXiv190600502F, 2019arXiv190407732F,    2019ApJ...871..200F,  2019arXiv190513572F})  scenarios have been explored.  Therefore,  photons with energies higher than  5 - 10 GeV as detected before \citep{1994Natur.372..652H} and during the Fermi LAT era \citep[][and therein]{2009Sci...323.1688A, 2014Sci...343...42A, 2013ApJS..209...11A} could  be evidence of the inverse Compton (IC) scattering existence. Several authors have taken into account the  two crucial similarities found in the Fermi LAT light curves and have concluded that VHE emission has its origin in external shocks \citep{2009MNRAS.400L..75K, 2010MNRAS.409..226K, 2010MNRAS.403..926G, 2014MNRAS.443.3578N, 2016ApJ...831...22F, 2009MNRAS.396.1163Z}. In particular, \cite{2013ApJ...771L..33W} showed that 10 - 100 GeV photons detected after the prompt phase could have originated by SSC emission of the early afterglow.    In the context of external shocks and requiring observations in other wavelengths,  several LAT-detected bursts have been described reaching similar conclusions \citep{2013ApJ...773L..20L, 2015MNRAS.454.1073B, 2016ApJ...818..190F}.\\
\\
For the first time, an excess of gamma-ray events with a significance of $\>$20 $\sigma$ during the first 20 minutes and photons with  energies  higher than 300 GeV was recently reported by the MAGIC collaboration from GRB 190114C \citep{2019GCN.23701....1M}.   This burst triggered the Burst Area Telescope (BAT) instrument  onboard Swift satellite  at 2019 January 14 20:57:06.012 UTC (trigger 883832) \citep{2019GCN.23688....1G} and it was followed up by  the Gamma-Ray Burst Monitor \citep[GBM;][]{2019GCN.23709....1D}, by LAT \citep{2019GCN.23709....1D}, by the X-ray Telescope \citep[XRT;][]{2019GCN.23688....1G, 2019GCN.23704....1O},  by the Ultraviolet/Optical Telescope \citep[UVOT;][]{2019GCN.23688....1G, 2019GCN.23725....1S}, by the SPI-ACS instrument \citep{2019GCN.23714....1M},  by the Mini-CALorimeter  instrument \citep{2019GCN.23712....1U}, by the Hard X-ray Modulation Telescope instrument \citep{2019GCN.23716....1X},  by Konus-Wind \citep{2019GCN.23737....1F}, by the Atacama Large Millimeter/submillimeter Array (ALMA),  by the Very Large Array (VLA) \citep{2019arXiv190407261L} and by several optical telescopes   \citep{2019GCN.23690....1T, 2019GCN.23693....1L, 2019GCN.23695....1S, 2019GCN.23699....1L, 2019GCN.23701....1M, 2019GCN.23702....1B, 2019GCN.23717....1I, 2019GCN.23726....1K, 2019GCN.23729....1D,  2019GCN.23740....1I, 2019GCN.23741....1M}.\\
\\
\cite{2019arXiv190201861R}  analyzed the GBM data  and found a typical prompt emission for the first $\sim$ 4 s, a smoothly broken-power law spectrum.  However, the GBM data for $\gtrsim$ 4 s showed  that i) the spectral evolution was consistent with a single component  similar to that of the LAT spectrum and ii)  the time of the bright peak coincided with the peak exhibited in the LAT data.    They concluded that both emissions were originated during the  afterglow phase.  Similarly,  \cite{2019arXiv190107505W} analyzed the  GBM and LAT data, finding that the MeV and GeV emission of GRB 190114C had the same origin during the afterglow evolution.  Using  the standard SSC model in a homogeneous medium,   \cite{2019arXiv190511312W} described the broadband SED of GRB 190114C during the first 150 s after the trigger time. These authors  concluded that the detection of the energetic  photons at hundred of GeVs was due to  the large burst energy and low redshift.   \cite{2019arXiv190508285D} argued that these photons were produced by the Comptonization of X-ray photons.\\
%
%
\cite{2019ApJ...879L..26F} analyzed  the gamma-ray (LAT and GBM), X-ray (BAT and XRT), optical (several telescopes) and radio (ALMA) light curves  of GRB 190114C. These authors showed that  the multi-wavelength observations during the first $\sim$ 400 s   were consistent with the external shock model evolving in a stratified stellar-wind like medium and after this time were consistent with a uniform ISM-like medium.     They also reported the external shock parameters they found using the Markov-chain Monte Carlo method when modelling the multi-wavelength (from radio to Fermi LAT) data.   Moreover, these authors argued  that the high-energy photons were  produced in the deceleration phase and that an alternative mechanism originated in  the forward shocks  should be considered to properly describe the energetic photons with energies beyond the synchrotron limit.      In particular,  the specific model that transitions from stratified stellar-wind to an homogeneous interstellar medium was chosen because the synchrotron seed photons for Comptonization  can reproduce the multi-wavelength observations, and also the VHE photons detected for almost 20 minutes by the MAGIC telescope, which covered the time lapse before and after this transition. It is worth noting that before this transition, as suggested by some authors \citep[e.g., see][]{2019arXiv190201861R, 2019arXiv190107505W, 2019ApJ...879L..26F}, the LAT, GBM, X-ray and optical observations are consistent with the evolution of the wind medium afterglow model, and after this transition the X-ray and optical observations with the constant medium afterglow model \citep[e.g., see][]{2019arXiv190511312W, 2019ApJ...879L..26F}.   Motivated by these results,  we extend the results shown in \cite{2019ApJ...879L..26F} and derive, in this paper, the SSC light curves and spectra in a  stratified stellar-wind medium, which transitions to an homogeneous interstellar medium.   The paper is organized as follows. In Section 2 we show  SSC light curves generated in the forward shock when the outflow decelerates in a stratified stellar-wind and homogeneous ISM-like medium.  In Section 3 we apply the SSC model to estimate the VHE emission of GRB 190114C using the parameters reported in \cite{2019ApJ...879L..26F}  and also discuss the results. In Section 4, conclusions are presented. \\ 
\vspace{0.8cm}
\section{SSC scenario of forward shocks} 
It is widely accepted that the standard synchrotron forward-shock model has been successful in describing the multiwavelength (X-ray, optical and radio) observations in GRB afterglows. However,  relativistic electrons are also expected to be cooled down by SSC emission \citep[e.g.][]{2001ApJ...548..787S}.   We do not discuss the effects of the self-absorption frequency, since it is typically relevant at low energies compared to the GeV  energy range \citep[e.g., see][]{2014ApJ...788...70P}. We do not use the reverse-shock emission because it was used in \cite{2019ApJ...879L..26F}  to explain the short-lasting Fermi LAT and GBM peaks at  $\sim 8\, {\rm s}$ and it cannot describe an emission much longer than this timescale.  Due to the absence of neutrinos spatially or temporally associated with GRB 190114C \citep{2019ATel12395....1V},  we neglect more complex models like hadronic or photo-hadronic processes \citep{2009ApJ...705L.191A, 2014MNRAS.437.2187F}.  They are by no means disfavored by these arguments.\\
The SSC forward-shock model varies the temporal and spectral features of GRB afterglows significantly and can also explain the gamma-rays above the well-known synchrotron limit \citep[e.g.,][]{2010ApJ...718L..63P, 2009ApJ...706L.138A, 2011MNRAS.412..522B, 2019ApJ...879L..26F}.   The SSC emission of a decelerating outflow moving through either a stratified or homogeneous medium is calculated in the next section.
\subsection{SSC light curves in the stratified stellar-wind medium}
When the outflow interacts with the stratified medium with density $\rho(r)=A\, r^{-2}$, where  $A=\frac{\dot{M}}{4\pi \rm v_W}$,\footnote{$\dot{M}$ is the mass-loss rate and v is the velocity of the outflow}  the minimum and the cooling  electron Lorentz factors  can be written as 
{\small
\bary\label{ele_Lorentz_w}
\gamma_{\rm m}&=&5.3\times 10^2\,g(p)\,\varepsilon_{e,-2}\,\Gamma_{2.5}\,, \cr
\gamma_{\rm c}&=&2.1\times 10^5 \left(\frac{1+Y_{\rm Th}}{3} \right)^{-1}\,\xi^2\,\varepsilon^{-1}_{B,-4}\,A^{-2}_{W,-1}\,\Gamma^{-3}_{2.5}\,E_{53.5}\,,
\eary
}
respectively. Hereafter, we adopt the convention $Q_{\rm x}=Q/10^{\rm x}$  in c.g.s. units.   The microphysical parameters $\varepsilon_{\rm B}$ and $\varepsilon_{\rm e}$ correspond to  the fraction of the shocked energy density transferred to the magnetic field and electrons, respectively,   the equivalent kinetic energy $E$ is associated with the isotropic energy $E_{\rm \gamma, iso}$ and the kinetic efficiency $\eta=E_{\rm \gamma, iso}/E$ which is defined as the fraction of the kinetic energy radiated into gamma-rays, $Y_{\rm Th}$ is the Compton parameter \citep{2001ApJ...548..787S, 2010ApJ...712.1232W}, $\xi$ is a constant parameter of order of unity \citep{2000ApJ...536..195C},  $g(p)=\frac{p-2}{p-1}\simeq0.23$ for $p=2.3$,   $\Gamma$ is the bulk Lorentz factor and $A_{\rm W}=A/(5\times 10^{11}\, {\rm g\, cm^{-1}})$ is the parameter of wind density \citep{2000ApJ...543...66P, 2000A&A...362..295V, 2005A&A...442..587V, 2004ApJ...606..369C, 1998MNRAS.298...87D, 2000ApJ...536..195C}.\\
Given the hydrodynamic forward-shock evolution in the stratified medium  {\small $\Gamma\sim 258.3\left(\frac{1+z}{1.5} \right)^{\frac14}E^{\frac14}_{53.5}A^{-\frac14}_{W, -1}t^{-\frac14}_2$} and the photon energy radiated by synchrotron  process $\epsilon_{\rm \gamma}(\gamma_e)\propto \Gamma B' \gamma^2_{\rm e}$  with $B'$ the comoving magnetic field,  the synchrotron spectral breaks and the maximum flux evolve as $\epsilon^{\rm syn}_{\rm m}\propto t^{-\frac32}$,  $\epsilon^{\rm syn}_{\rm c}\propto t^{\frac12}$ and $F^{\rm syn}_{\rm max}\propto t^{-\frac12}$, respectively \citep[][]{2000ApJ...543...66P}. \\
Photons generated by synchrotron radiation can be up-scattered in the forward shocks by the same electron population as $\epsilon^{\rm ssc}_{\rm m (c)}\sim\gamma^2_{\rm m(c)} \epsilon^{\rm syn}_{\rm m(c)}$ with a maximum flux given by $F^{\rm ssc}_{\rm max}\sim\,k\tau\,F^{\rm syn}_{\rm max}$, where  {\small $\tau$} is the optical depth and {\small $k=\frac{4(p-1)}{p-2}$}. Therefore, taking into account the electron Lorentz factors (eq. \ref{ele_Lorentz_w}) and the synchrotron spectral breaks \citep{2000ApJ...536..195C},  the SSC spectral breaks  and the maximum flux for SSC emission can be written as 
{\small
\bary\label{ssc-br_w}
\epsilon^{\rm ssc}_{\rm m}&\simeq&15.1\,{\rm keV} g(p)^4\,  \left(\frac{1+z}{1.5} \right)\,\xi^{\frac12} \,\varepsilon_{e,-2}^{4}\,\varepsilon_{B,-4}^{\frac12}\,A^{-\frac12}_{W,-1}\,E_{53.5}\,t^{-2}_2\,,\cr
\epsilon^{\rm ssc}_{\rm c}&\simeq& 97.5\,{\rm TeV} \, \left(\frac{1+z}{1.5} \right)^{-3}\,\left(\frac{1+Y_{\rm Th}}{3} \right)^{-4}\, \xi^{\frac92} \,\varepsilon_{B,-4}^{-\frac72}\,A^{-\frac92}_{W,-1}\cr
&&\hspace{5.6cm}\times\,E_{53.5}\,t^{2}_2\,,\cr
F^{\rm ssc}_{\rm max}&\simeq& 1.1\times 10^{-4}\,{\rm mJy} \,\left(\frac{1+z}{1.5} \right)^{2}\,\xi^{-\frac52}\,\varepsilon_{B,-4}^{\frac12}\,A^{\frac52}_{W}\,D^{-2}_{27}\, t^{-1}_2\,,
\eary
}
where $z=0.5$ is the redshift and $D$ is the luminosity distance of the burst. The luminosity distance is obtained using the values of cosmological parameters reported in \cite{2018arXiv180706209P}: the matter density parameter of $\Omega_{\rm m}=0.315\pm 0.007$ and the Hubble constant of $H_0=67.4\pm 0.5\,{\rm km\,s^{-1}\,Mpc^{-1}}$.\\
During the deceleration phase the intrinsic attenuation by $e^\pm$ pair production due to collision of a VHE photon with a lower-energy photon is given by \cite[e.g., see][]{2009grb..book.....V}
\be
\tau_{\rm \gamma\gamma,in}\simeq 10^{-2}\,\, R_{\rm dec, 17}\,\Gamma^{-1}_{2.5}\, n_{\rm \gamma,10.8}\,,
\ee
where  $R_{\rm dec}=1.17\times10^{17}\,{\rm cm} \left(\frac{1+z}{1.5} \right)^{\frac12}\,\xi \, A^{-\frac12}_{-1}\,E^{\frac12}_{53.5}\,t^{\frac12}_2$ is the deceleration radius and  $n_\gamma\simeq 6\times 10^{10}\,{\rm cm^{-3}}\, L_{\rm \gamma, 50}\,R^{-2}_{\rm dec, 17}\,\Gamma^{-1}_{2.5}\,\epsilon^{-1}_{\rm \gamma,3}$ is the keV-photon density ($\epsilon_{\rm \gamma,3}=$1 keV) with $L_{\rm \gamma}$ the keV-photon luminosity. Since  $\tau_{\rm \gamma\gamma,in}\ll 1$  during the deceleration phase,  the intrinsic attenuation (opacity) is not considered.\\ 
Given the SSC spectra for fast- and slow-cooling regime  together with the SSC spectral breaks and the maximum flux (eq. \ref{ssc-br_w}),  the SSC light curves in the fast (slow)-cooling regime are   
%
%
%
{\small
\begin{eqnarray}
\label{ssc_wind}
F^{\rm ssc}_{\nu}= \cases{ 
F^{\rm W}_{\rm 1,f(s)}t^{-\frac{5}{3}}_2\,(t^{-\frac13}_2) \epsilon_{\gamma,11}^{\frac13},\hspace{0.69cm} \epsilon_\gamma < \epsilon^{\rm ssc}_{\rm c}(\epsilon^{\rm ssc}_{\rm m}), \cr
F^{\rm W}_{\rm 2,f(s)}t^{0}_2 \epsilon_{\gamma,11}^{-\frac12}(t^{-p}_2 \epsilon_{\gamma,11}^{-\frac{p-1}{2}}),\,\,\,\, \hspace{0.0cm}\epsilon^{\rm ssc}_{\rm c}(\epsilon^{\rm ssc}_{\rm m})<\epsilon_\gamma<\epsilon^{\rm ssc}_{\rm m} (\epsilon^{\rm ssc}_{\rm c}),\hspace{.3cm} \cr
F^{\rm W}_{\rm 3,f(s)}t^{1-p}_2\,\epsilon_{\gamma,11}^{-\frac{p}{2}},\hspace{1.35cm} \{\epsilon^{\rm ssc}_{\rm m},\epsilon^{\rm ssc}_{\rm c}\}<\epsilon_\gamma\,, \cr
}
\end{eqnarray}
}
where $\epsilon_{\gamma,11}=100\,{\rm GeV}$ and $t_2=100\,{s}$ correspond to the energy band and timescale at which the flux is estimated.  The values of the proportionality constants $F^{\rm W}_{\rm m,n}$ for $m=1$, $2$ and $3$ and $n=f$ (fast) or $s$ (slow) are reported in appendix A.  The  SSC light curves agree with the ones derived in \cite{2000ApJ...543...66P} for a stratified medium. It is worth noting that these authors calculated the light curves for the energy band of X-rays and timescales of days. \\
In the SSC spectrum,  the Klein-Nishina (KN) regime must be considered because the emissivity beyond this frequency is drastically decreased compared with the classical Thomson regime. The spectral break caused by the decrease of the scattering cross section, due to the  KN effects, is given by 
{
\small
\bary\label{kn}
\epsilon^{\rm ssc}_{\rm KN}&\simeq& 8.8\,{\rm TeV}\,\left(\frac{1+z}{1.5} \right)^{-\frac32} \left(\frac{1+Y_{\rm Th}}{3} \right)^{-1}\xi^2\,\varepsilon_{\rm B, -4}^{-1}\,A^{-\frac32}_{\rm W,-1}\,\cr
&&\hspace{5.4cm} \times\, E^{\frac12}_{53.5}\,t^{\frac12}_2.
\eary
}
The SSC light curves given in eq.~(\ref{ssc_wind}) show two features:   i) The second PL segment of the fast-cooling regime ($\epsilon^{\rm ssc}_{\rm c}<\epsilon_\gamma<\epsilon^{\rm ssc}_{\rm m}$) does not evolve with time, and others decrease gradually.    It indicates that SSC emission is more likely to be detected during the first seconds after the trigger time, and  ii) The first PL segment in the fast-cooling regime ($\epsilon_\gamma<\epsilon^{\rm ssc}_{\rm c}$), and the third PL segments ($\{\epsilon^{\rm ssc}_{\rm m},\epsilon^{\rm ssc}_{\rm c}\}<\epsilon_\gamma$) are decaying functions of the circumburst density.    It suggests that depending on the timescale and energy range observed, the SSC emission could be detected in environments with higher and/or lower densities.\\
The top panels in Figure \ref{fig1:wind} show the resulting light curves and SEDs of the SSC forward-shock emission generated by a decelerating outflow in a stratified medium. These panels were obtained using relevant values for GRB afterglows.\footnote{$E=5\times10^{53}\,{\rm erg}$, $p=2.3$,  $\varepsilon_{\rm B}=10^{-3}$, $\varepsilon_{\rm e}=10^{-2}$, $z=0.5$, $\xi=1$ and $A_{W}=10^{-3}\,(10^{-1})$ for the purple (green) curve.}   The observable quantities, the microphysical parameters, the parameter density and the efficiencies  are in the range proposed to produce GeV photons in the afterglow phase \citep[e.g., see][]{2015MNRAS.454.1073B, 2016MNRAS.461...51B}. The effect of the extragalactic background light (EBL) absorption proposed by \cite{2017A&A...603A..34F} was used.  The gold and blue solid curves in the left-hand panel correspond to 10 and 100 GeV, respectively, and the gold and blue dashed-dotted curves to the Fermi LAT \citep{2016CRPhy..17..617P}  and the MAGIC \citep{2008ApJ...687L...5T} sensitivities at the same energies, respectively. The purple and green curves in the  top right-hand panel correspond to the SEDs at 10 and 100 s, respectively.  \\
The top panels show that the SCC flux is very sensitive to the external density.  The light curves above the LAT and MAGIC sensitivities are obtained with $A_{\rm W}=10^{-1}$ and below with $A_{\rm W}=10^{-3}$.   In this particular case, both light curves (at 10 and 100 GeV) evolve in the second PL segment of slow-cooling regime.  The spectral breaks are $\epsilon^{\rm ssc}_{\rm m}=0.5\,{\rm MeV}$ and $\epsilon^{\rm ssc}_{\rm  c}=2.8\times 10^8\,{\rm TeV}$ for $A_{\rm W}=10^{-3}$, and $\epsilon^{\rm ssc}_{\rm m}=47.5\,{\rm keV}$ and $\epsilon^{\rm ssc}_{\rm  c}=0.2\,{\rm TeV}$ for $A_{\rm W}=0.1$.  The transition times between fast to slow cooling regime are $0.03$ and $0.3$ s for $A_{W}=10^{-3}$ and $0.1$, respectively. It shows that with these parameters the SSC emission decreases monotonically with time and  increases as the density of the circumburst medium increases. For the chosen parameters, the break energies in the KN regime are $1.2\times 10^3\,{\rm TeV}$ and  $1.3\,{\rm TeV}$ for $A_{W}=10^{-3}$ and $0.1$, respectively, which are above the energies of the Fermi and MAGIC sensitivities.   We emphasize that depending on the parameter values,  the SSC emission would lie in the KN regime, and then this emission would be drastically suppressed. Similarly,  the electron distribution that up-scatters synchrotron forward-shock photons beyond the KN regime would be affected \citep{2009ApJ...703..675N, 2010ApJ...712.1232W} and also the degree of cooling of synchrotron emitting electrons would be affected by KN \citep{2015MNRAS.454.1073B}.\\
The top left-hand panel of Fig. \ref{fig1:wind} shows that the SCC flux is above the Fermi LAT and MAGIC sensitivities during the first 100 s for $A_{W}=0.1$  but not for  the value of $A_{W}=10^{-3}$.   Therefore, the probability to observe the SSC emission from the GRB afterglow is higher during the first seconds after the burst trigger  than at late times, and when the stellar wind ejected by the progenitor is denser.\\ 
%
%
The top right-hand panel shows the SEDs for the same set of parameters at $t=10$ and 100 s.    The value of $A_{W}=0.1$ corresponds to the curve above the Fermi LAT and MAGIC sensitivities and the value of $A_{\rm W}=10^{-3}$ to the curve below the sensitivities.  The red dashed line corresponds to the maximum energy radiated by synchrotron.  The filled areas in gray and cyan colors correspond to the Fermi LAT and MAGIC energy ranges, respectively.  The Fermi LAT and MAGIC areas show that photons above the synchrotron limit can be explained by SSC emission.  In addition, the top right-hand panel displays that the maximum SSC flux, due to EBL absorption, lies at i) the end of the LAT  energy range where this instrument has less sensitivity \citep{FUNK2013348} and ii) the beginning of the MAGIC energy range, making this telescope ideal for detecting the SSC emission.\\
In order to compare the synchrotron and SSC fluxes at Fermi LAT energies (e.g. $\epsilon_\gamma=800\,{\rm MeV}$), we obtain the synchrotron and SSC spectral breaks   at $t=10^2\,{\rm s}$ for $A_{W}=0.1$ ($\epsilon^{\rm syn}_{\rm m}=0.2\,{\rm eV}$, $\epsilon^{\rm syn}_{\rm  c}=0.3\,{\rm keV}$  $\epsilon^{\rm ssc}_{\rm m}=47.5\,{\rm keV}$ $\epsilon^{\rm ssc}_{\rm  c}=0.2\,{\rm TeV}$).  Therefore, at the Fermi LAT energy range the synchrotron emission evolves in the third PL segment of slow-cooling regime and SSC emission in the second PL segment. In this case,  the ratio of synchrotron and SCC fluxes becomes
{\small
\bary\label{wind}
\frac{F^{\rm syn}_\nu}{F^{\rm ssc}_\nu} &\sim&  1.6\left(\frac{1+z}{1.5} \right)^{\frac14}\,\xi^{\frac72}\left(\frac{1+Y_{\rm Th}}{ 2} \right)^{-1}\, \varepsilon^{-p+\frac14}_{e,-2}A^{-\frac52}_{W,-1}\, E^{\frac34}_{53.5}\cr
&& \hspace{3.2cm} \,\Gamma^{1-p}_{2.5}\,t^{\frac34}_2\,\left( \frac{\epsilon_\gamma}{800\,{\rm MeV}}\right)^{-\frac12}\,,
\eary
}
which is  of order unity.  Here, we use a smaller value of the effective Compton Y parameter because electrons radiating synchrotron at these large energies have a Compton Y parameter smaller than the corresponding value in the Thomson regime. \footnote{The energy break of scattering photons above which the scatterings with the electron population given by Lorentz factor $\gamma_{\rm e}$ lie the KN  regime is given by $\epsilon^{\rm syn}_{\rm KN}(\gamma_{\rm e})=m_{\rm e}\,c^2\, \frac{\Gamma}{\gamma_{\rm e}}$  \citep{2010ApJ...712.1232W,  2015MNRAS.454.1073B}.  For $\gamma_{\rm e}=\gamma_{\rm c}$,  eqs. \ref{ele_Lorentz_w} and \ref{kn} are related by $\epsilon^{\rm syn}_{\rm KN}(\gamma_{\rm c})=  \frac{\epsilon^{\rm ssc}_{\rm KN}}{\gamma^2_{\rm c}}\simeq2.1\,{\rm keV}$. Taking into account that the electron Lorentz factor of $\gamma^*_e\simeq 3\times 10^7$ produces the synchrotron photons at $\sim$ 300 MeV, the corresponding KN photon energy is $\epsilon^{\rm syn}_{\rm KN}(\gamma^*_{\rm e})\simeq 1.4\,{\rm eV}$. Given that the characteristic and cutoff synchrotron breaks are $\epsilon^{\rm syn}_{\rm m}=1.6\,{\rm eV}$ and $\epsilon^{\rm syn}_{\rm c}=0.2\,{\rm keV}$, respectively, the Compton parameter lies in the range $\epsilon^{\rm syn}_{\rm KN}(\gamma^*_{\rm e})<\epsilon^{\rm syn}_{\rm m}< \epsilon^{\rm syn}_{\rm c} < \epsilon^{\rm syn}_{\rm KN}(\gamma_{\rm c})$. For this case, $Y(\gamma^*_{\rm e})=Y_{\rm Th}\,\left(\frac{\epsilon^{\rm syn}_{\rm m}}{\epsilon^{\rm syn}_{\rm c}}\right)^{\frac{3-p}{2}}\,\left(\frac{\epsilon_{\rm KN}(\gamma^*_{\rm e})}{\epsilon^{\rm syn}_{\rm m}}\right)^\frac43\simeq 0.57\, Y_{\rm Th}$ with $Y_{\rm Th}=1.93$.}  We can conclude that below $\sim$ 300 MeV the flux can be described in the synchrotron forward-shock scenario,  between $\sim$ 0.5 - 1 GeV the contribution of  both processes would be relevant,  and  beyond the synchrotron limit,  the observations would be entirely explained by SSC process for this set of parameters. \\
%
%
The top right-hand panel shows that the maximum SSC flux  lies at 100 GeV,  making  it possible to detect the VHE emission in observatories where the sensitivity is maximum  at hundreds of GeV (e.g. MAGIC telescope) but not in those observatories where the maximum sensitivity lies in few TeVs \citep[e.g. High Altitude Water Cherenkov (HAWC);][]{2012APh....35..641A}.  For instance,  the SSC flux at 1 TeV decreases between two and three orders of magnitude in comparison with the flux at 100 GeV.\\
Given the minimum and cooling electron Lorentz factors (eq. \ref{ele_Lorentz_w}), the synchrotron and SSC luminosity ratio can be computed as \footnote{This relation was obtained using the synchrotron and SSC luminosity ratio derived in \cite{2001ApJ...548..787S}  with the equivalent density for the stratified medium.}  
\be
\frac{L^{\rm ssc}_\nu}{L^{\rm syn}_\nu}\simeq 0.41\, A_{\rm W,-1}\, R^{-1}_{\rm dec, 17}\gamma_{\rm c,5}^2\left(\frac{\gamma_{\rm c,5}}{\gamma_{\rm m,2}}\right)^{1-p}\,.
\ee
  It is worth mentioning that the synchrotron and SSC luminosity ratio depends on $Y$ through $\gamma_{\rm c}$.  Therefore, in the case of a stratified medium, half of the synchrotron luminosity is up-scattered by SSC emission.
\subsection{SSC light curves in a homogeneous ISM-like medium}
\begin{figure}[h!]
{ \centering
\resizebox*{0.5\textwidth}{0.5\textheight}
{\includegraphics{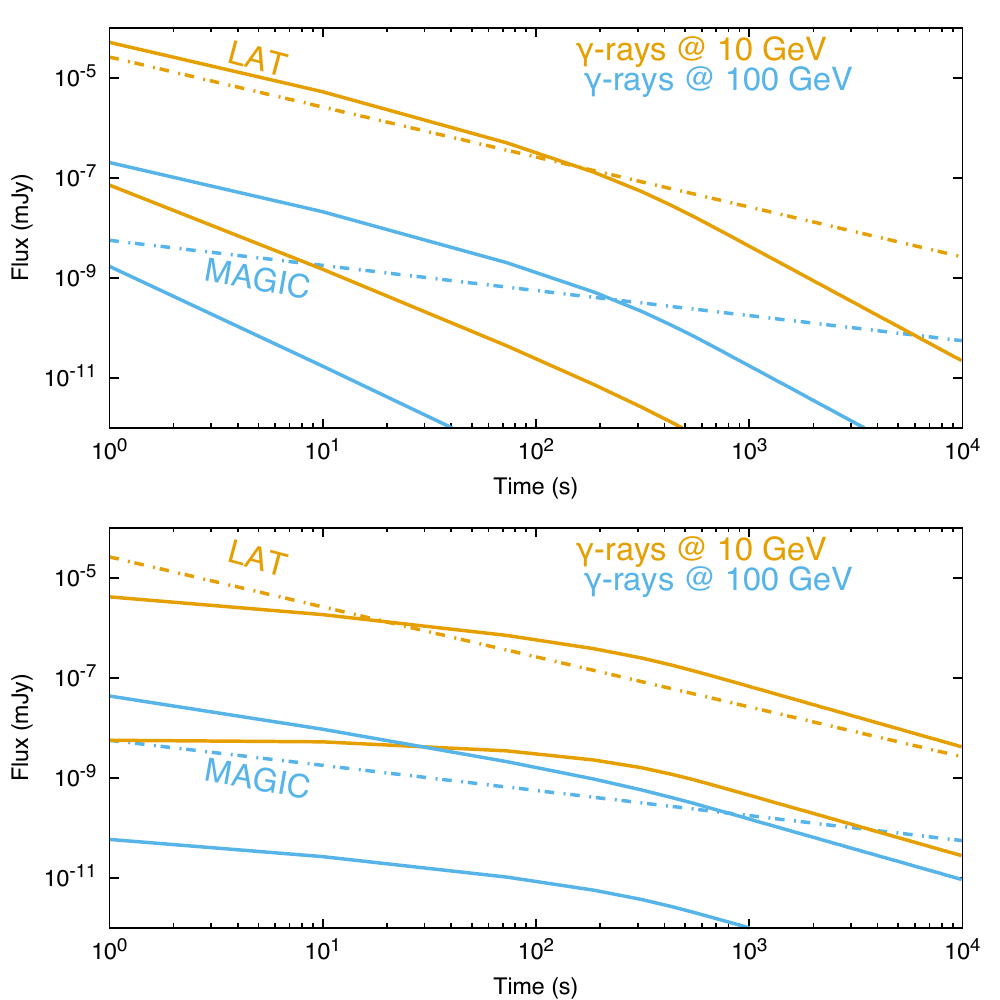}}
\resizebox*{0.5\textwidth}{0.5\textheight}
{\includegraphics{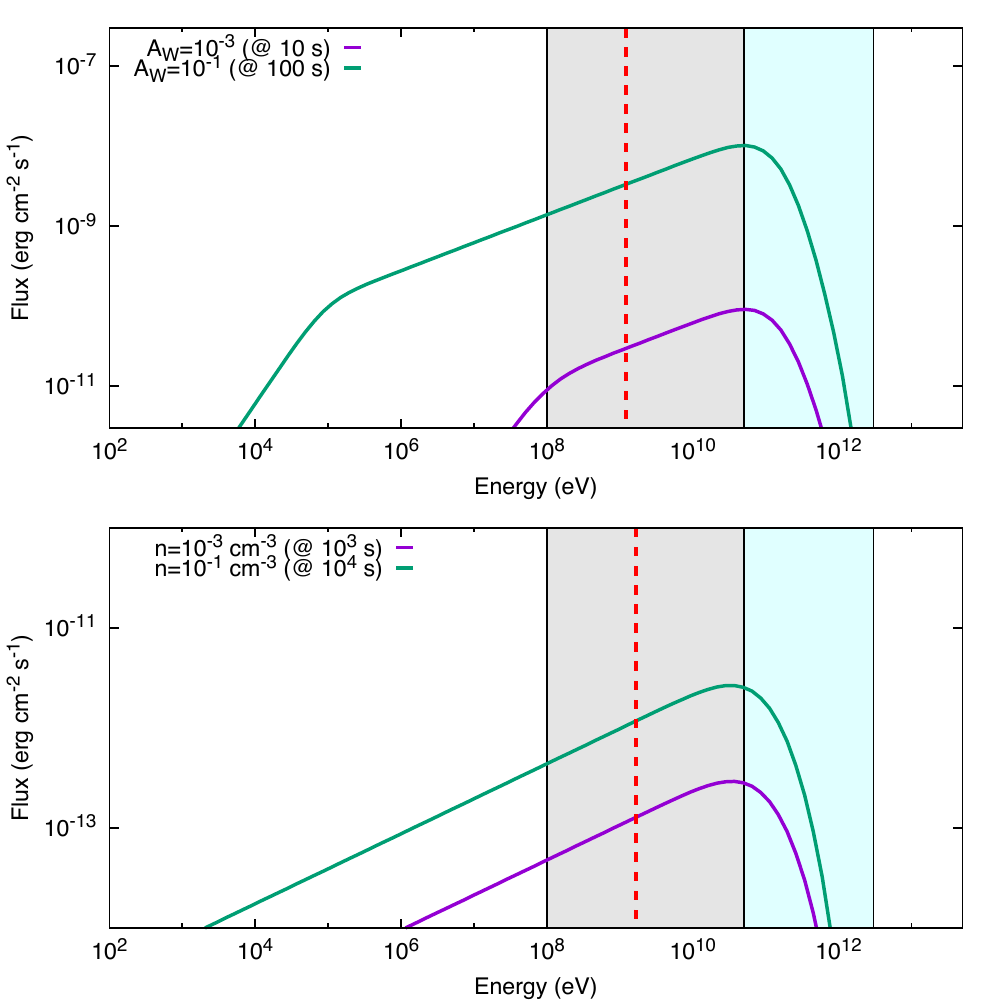}}
}
\caption{The figures show the light curves (left) and SEDs (right) of SSC emission expected from a  decelerating outflow in a stratified (above) and homogeneous (below) medium  for values of $A_{W}=10^{-3}\,(10^{-1})$ and $n=10^{-3}\,(10^{-1})\,{\rm cm^{-3}}$, respectively.  In all panels  the values of $E=5\times 10^{53}\,{\rm erg}$, $p=2.3$,  $\varepsilon_{\rm B}=10^{-3}$, $\varepsilon_{\rm e}=10^{-2}$ and $z=0.5$ were used.  Dashed-dotted lines in left panels represent the LAT \citep[gold; ][]{2016CRPhy..17..617P} and MAGIC \citep[blue;][]{2008ApJ...687L...5T} sensitivities at 10 and 100 GeV respectively.  The light curves above the LAT and MAGIC sensitivities are obtained with $A_{\rm W}=10^{-1}$ and $n=10^{-1}\,{\rm cm^{-3}}$ for a stratified and homogeneous medium  and below the sensitivities are obtained with $A_{\rm W}=10^{-3}$ and $n=10^{-3}\,{\rm cm^{-3}}$, respectively.  The SEDs are shown at two different times, $t=10$ and $100$ s for the wind medium, and $t=10^3$ and $10^4$ s for the homogeneous medium.  The red dashed lines correspond to the synchrotron limit.  The filled areas in gray and cyan colors correspond to the Fermi LAT and MAGIC energy ranges, respectively.  The effect of the extragalactic background light (EBL) absorption proposed by \cite{2017A&A...603A..34F}.}
 \label{fig1:wind}
\end{figure}

When the outflow interacts with a homogeneous medium with density $n$,  the minimum and the cooling electron Lorentz factors  can be written as 
{\small
\bary\label{elec_factor}
\gamma_{\rm m}&=& 4.3\times 10^2  \,g(p) \,\varepsilon_{\rm e} \,\Gamma_{2}, \cr
\gamma_{\rm c}&=& 1.9\times 10^4\, \left(\frac{1+Y_{\rm Th}}{1.5} \right)^{-1}\,\left(\frac{1+z}{2.5} \right)\,\varepsilon^{-1}_{B,-4}\,n^{-1}\,\Gamma^{-3}_{2}\,t^{-1}_3.\,\,\,\,\,\,\,
\eary
}
Given the hydrodynamic forward-shock evolution in the homogeneous medium  {\small $\Gamma\sim 101.2\,  \left(\frac{1+z}{1.5} \right)^{\frac38}E^{\frac18}_{53.5}n^{-\frac18}t^{-\frac38}_3$}  and the photon energy radiated by synchrotron  $\epsilon_{\rm \gamma}(\gamma_e)\propto \Gamma B' \gamma^2_{\rm e}$,  the synchrotron spectral breaks and the maximum flux evolve as  $\epsilon^{\rm syn}_{\rm m}\propto t^{-\frac32}$, $\epsilon^{\rm syn}_{\rm c}\propto t^{-\frac12}$ and $F^{\rm syn}_{\rm max}\propto t^0$ \citep[e.g.,][]{1998ApJ...497L..17S}. \\
Taking into consideration the electron Lorentz factors (eq. \ref{elec_factor}) and the synchrotron spectral breaks \citep{1998ApJ...497L..17S},  the spectral breaks  and the maximum flux for SSC emission can be written as 
{\small
\bary\label{ssc_br-h}
\epsilon^{\rm ssc}_{\rm m}&\simeq& 11.4\,{\rm keV}\,g(p)^4\,  \left(\frac{1+z}{1.5} \right)^{\frac54}\,\varepsilon_{e,-2}^{4}\,\varepsilon_{B,-4}^{\frac12}\,n^{-\frac14}\,E^{\frac34}_{53.5}\,t^{-\frac94}_3,\cr
\epsilon^{\rm ssc}_{\rm c}&\simeq& 0.1\,{\rm TeV} \, \left(\frac{1+z}{1.5} \right)^{-\frac34}\,\left(\frac{1+Y_{\rm Th}}{2.5} \right)^{-4}\,\varepsilon_{B,-4}^{-\frac72}\,n^{-\frac94}\,E^{-\frac54}_{53.5},\cr
&&\hspace{6.2cm}\times\,t^{-\frac14}_3\,,\cr
F^{\rm ssc}_{\rm max}&\simeq& 1.1\times 10^{-5}\,{\rm mJy} \,\left(\frac{1+z}{1.5} \right)^{\frac34}\,\varepsilon_{B,-4}^{\frac12}\,n^{\frac54}\,D^{-2}_{27}\,E^{\frac54}_{53.5}\, t^{\frac14}_3.
\eary
}
Given the synchrotron spectra for fast- and slow-cooling regime  together with the SSC spectral breaks and the maximum flux (eq. \ref{ssc_br-h}),  the SSC light curves in the fast (slow)-cooling regime  are \citep{2001ApJ...548..787S}
{\footnotesize
\begin{eqnarray}
\label{ssc_ism}
F^{\rm ssc}_{\nu}= \cases{ 
F^{\rm H}_{\rm 1,f(s)}t^{\frac{1}{3}}_3\, (t_3)  \epsilon_{\gamma,11}^{\frac13},\hspace{2.cm} \epsilon_\gamma<\epsilon^{\rm ssc}_{\rm c} (\epsilon^{\rm ssc}_{\rm m}), \cr
F^{\rm H}_{\rm 2,f(s)}t^{\frac{1}{8}}_3 \epsilon_{\gamma,11}^{-\frac12}(t^{-\frac{9p-11}{8}}_3 \epsilon_{\gamma,11}^{-\frac{p-1}{2}}), \hspace{0.cm} \epsilon^{\rm ssc}_{\rm \gamma,c} (\epsilon^{\rm ssc}_{\rm m})<\epsilon_\gamma<\epsilon^{\rm ssc}_{\rm m} (\epsilon^{\rm ssc}_{\rm c}),\hspace{.25cm}\cr
F^{\rm H}_{\rm 3,f(s)}t^{-\frac{9p-10}{8}}_3\,\epsilon_{\gamma,11}^{-\frac{p}{2}},\,\,\,\, \hspace{1.5cm}  \{\epsilon^{\rm ssc}_{\rm c},\epsilon^{\rm ssc}_{\rm m}\}<\epsilon_\gamma\,, \cr
}
\end{eqnarray}
}
where $\epsilon_{\gamma,11}=100\,{\rm GeV}$ and $t_3=1000\,{s}$ correspond to the energy band and timescale at which the flux is estimated.  The values of the proportionality constants $F^{\rm H}_{\rm m,n}$ for $m=1$, $2$ and $3$ and $n=f$ (fast) or $s$ (slow) are reported in appendix A.  It is worth noting that the time evolution of each PL segment of the SSC light curve agrees with the ones derived in  \cite{2000ApJ...543...66P} for the fast- and slow-cooling regime  and   the PL segments in the slow-cooling regime derived by \cite{2001ApJ...548..787S}.\\
 For the case of the homogeneous medium, the spectral break caused by the decrease of the scattering cross section, due to the  KN effects, is given by
{
\small
\bary
\epsilon^{\rm KN}_{\rm c}&\simeq& 0.7\,{\rm GeV}\, \left(\frac{1+z}{1.5} \right)^{-\frac34}\, \left(\frac{1+Y_{\rm Th}}{2.5} \right)^{-1} \,\varepsilon_{B,-4}^{-1}\,n^{-\frac34}\,E^{-\frac14}_{53.5}\,\cr
&&\hspace{5.9cm}\times\, t^{-\frac14}_3.
\eary
}
The light curves given in eq.~(\ref{ssc_ism}) show two important features:  i) They show that at lower energies the SSC flux increases with time and at higher energies it decreases in both regimes (fast and slow). Therefore, it indicates that SSC emission is more probable to be detected during the first seconds after the trigger time,  although if this emission is very strong it can be observed for long times.  It is worth  highlighting that the SSC emission at lower energies  is eclipsed by the synchrotron radiation, and ii) The first PL segment in the fast-cooling regime  ($\epsilon_\gamma<\epsilon^{\rm ssc}_{\rm c}$), and the third PL segments  ($\{\epsilon^{\rm ssc}_{\rm c},\epsilon^{\rm ssc}_{\rm m}\}<\epsilon_\gamma$) are decaying functions of the circumburst density.    It suggests that depending on the timescale and energy range observed, the SSC emission could be detected in environments with higher and/or lower densities.\\
The  bottom panels in Figure \ref{fig1:wind}  show the resulting light curves and SEDs of the SSC forward-shock emission generated by  a decelerating outflow in  a homogeneous medium.  These panels were obtained using relevant values for GRB afterglows.\footnote{$E=5\times10^{53}\,{\rm erg}$, $p=2.3$,  $\varepsilon_{\rm B}=10^{-3}$, $\varepsilon_{\rm e}=10^{-2}$, $z=0.5$ and $n=10^{-3}\,(10^{-1})\, {cm^{-3}}$ for the purple (green) curve.} The observable quantities, the microphysical parameters, the circumburst density and the efficiencies for a homogeneous density are in the range proposed to produce GeV photons in the afterglow phase \citep[e.g. see,][]{2015MNRAS.454.1073B, 2016MNRAS.461...51B}.   Again, the effect of the EBL absorption proposed by \cite{2017A&A...603A..34F} was considered.   The bottom left-hand panel shows that the SSC flux is above the Fermi LAT and MAGIC sensitivities during the first 100 s for $n=0.1\,{\rm cm^{-3}}$ but not for the value of $n=10^{-3}\,{\rm cm^{-3}}$.  The gold and blue solid curves in the top left-hand panel correspond to 10 and 100 GeV, respectively, and the gold and blue dashed-dotted curves to the Fermi LAT \citep{2016CRPhy..17..617P}  and the MAGIC \citep{2008ApJ...687L...5T} sensitivities at the same energies, respectively.  The purple and green curves in the right-hand panel correspond to the SEDs at 10 and 100 s, respectively.\\
The  bottom panels of Fig. \ref{fig1:wind} show that the SSC flux is very  sensitive to the external density.  The light curves above the LAT and MAGIC sensitivities are obtained with $n=10^{-1}\,{\rm cm^{-3}}$ and below with $n=10^{-3}\,{\rm cm^{-3}}$.    In this particular case, both light curves (at 10 and 100 GeV) evolve in the second PL segment of slow-cooling regime.  The spectral breaks are $\epsilon^{\rm ssc}_{\rm m}=36.1\,{\rm keV}$ and $\epsilon^{\rm ssc}_{\rm  c}=5.3\times 10^3\,{\rm TeV}$ for $n=10^{-3}\,{\rm cm^{-3}}$, and $\epsilon^{\rm ssc}_{\rm m}=11.4\,{\rm keV}$ and $\epsilon^{\rm ssc}_{\rm  c}=0.2\,{\rm TeV}$ for $n=0.1\,{\rm cm^{-3}}$, respectively.   The transition times between fast- to slow-cooling regime are $0.06$ and $0.6$ s for $n=10^{-3}\,{\rm cm^{-3}}$ and $0.1\,{\rm cm^{-3}}$, respectively.  It shows that with the chosen values the SSC emission decreases monotonically with time and  increases as  the density of the circumburst medium increases.  Using the parameter values, the break energies in the KN regime are $31.7\,{\rm TeV}$ and  $1.1\,{\rm TeV}$ for $n=10^{-3}$ and $0.1\,{\rm cm^{-3}}$, respectively, which are above the energies of the Fermi and MAGIC sensitivities.  Again,  we emphasize that depending on the parameter values,  the SSC emission would lie in the KN regime, and then this will be drastically suppressed. Similarly,  the electrons population that up-scatters synchrotron photons beyond the KN regime would be altered \citep{2009ApJ...703..675N, 2010ApJ...712.1232W}.\\
The  bottom left-hand panel of Fig. \ref{fig1:wind} shows that the SSC flux at 10 GeV is above the Fermi LAT sensitivity after $\sim$ 30 s and at 100 GeV is above MAGIC sensitivity during the first 850 s for $n=0.1\,{\rm cm^{-3}}$  but not for  $n=10^{-3}\,{\rm cm^{-3}}$.   Therefore, the probability to detect the SSC emission from the GRB afterglow depends on the observed energy.   For $\epsilon_{\gamma}=10\,{\rm GeV}$, the SSC emission could be detected delayed with respect to the prompt phase whereas  for  $\epsilon_{\gamma}=100\,{\rm GeV}$ it could be detected in temporal coincidence with lower-energy photons.\\ 
The  bottom right-hand panel of Fig. \ref{fig1:wind} shows the SEDs for the same parameter densities at $t=10^3$ and $10^4$ s. The value of $n=0.1\,{\rm cm^{-3}}$ corresponds to the curve above the Fermi LAT and MAGIC sensitivities and the value of $n=10^{-3}\,{\rm cm^{-3}}$ to the curve below these sensitivities. The red dashed line corresponds to the maximum energy radiated by synchrotron.  The filled areas in gray and cyan colors correspond to the Fermi LAT and MAGIC energy ranges, respectively.  The Fermi LAT and MAGIC areas show that photons above the synchrotron limit can be explained by SSC emission, similar to the case of the stratified medium.  The maximum SSC flux, due to EBL absorption, lies at the lower end of the MAGIC energy range, making this telescope ideal for detecting the SSC emission generated in a homogenous medium.\\
In order to compare the synchrotron and SSC fluxes at Fermi LAT energies (e.g. $\epsilon_\gamma=800\,{\rm MeV}$), we obtain the synchrotron and SSC spectral breaks  at $t=10^3\,{\rm s}$ for $n=0.1\,{\rm cm^{-3}}$ ($\epsilon^{\rm syn}_{\rm m}=6.1\times 10^{-2}\,{\rm eV}$, $\epsilon^{\rm syn}_{\rm  c}=0.3\,{\rm keV}$  $\epsilon^{\rm ssc}_{\rm m}=11.4\,{\rm keV}$ $\epsilon^{\rm ssc}_{\rm  c}=0.2\,{\rm TeV}$).  Therefore, at the Fermi LAT energy range the synchrotron emission evolves in the third PL segment of slow-cooling regime and the SSC emission in the second PL segment. In this case,  the ratio of synchrotron and SSC fluxes becomes
{\small
\bary\label{ism}
\frac{F^{\rm syn}_\nu}{F^{\rm ssc}_\nu} &\sim& 1.9 \left(\frac{1+z}{1.5} \right)^{-\frac14}\,\left(\frac{1+Y_{\rm Th}}{1.4} \right)^{-1}\, \varepsilon^{1-p}_{e,-2} \varepsilon^{-\frac34}_{B,-4}n^{-\frac76}_{-1}\, E^{-\frac{1}{12}}_{53.5}\cr
&& \hspace{3.cm} \,\Gamma^{\frac53-p}_{2.5}t^{-\frac14}_2\,\left( \frac{\epsilon_\gamma}{800\,{\rm MeV}}\right)^{-\frac12}\,,
\eary
}
which is  of order unity.    We want to emphasize  that the synchrotron and SSC flux ratio depends explicitly on $Y$.  For the homogeneous medium,  we use a smaller value of the effective Compton Y parameter because electrons radiating synchrotron at these large energies have a Compton Y parameter smaller than the corresponding value in the Thomson regime. \footnote{For the homogeneous medium,  an analysis of the effective Y parameter for electrons radiating synchrotron at Fermi LAT energies can also be done. In this case,  the break energies are $\epsilon_{\rm KN}(\gamma_{\rm c})=  \frac{\epsilon^{\rm ssc}_{\rm KN}}{\gamma^2_{\rm c}}\simeq0.9\,{\rm keV}$, $\epsilon_{\rm KN}(\gamma^*_{\rm e})\simeq 0.6\,{\rm eV}$,  $\epsilon^{\rm syn}_{\rm m}=0.7\,{\rm eV}$ and $\epsilon^{\rm syn}_{\rm c}=0.1\,{\rm keV}$. Again,  the Compton parameter corresponding to the case $\epsilon_{\rm KN}(\gamma^*_{\rm e})<\epsilon^{\rm syn}_{\rm m}< \epsilon^{\rm syn}_{\rm c} < \epsilon_{\rm KN}(\gamma_{\rm c})$ is $Y(\gamma^*_{\rm e})\simeq0.28\, Y_{\rm Th}$ with $Y_{\rm Th}=1.45$.  Therefore, a similar conclusion to that found in the stratified case is given.}  We can conclude that below $\sim$ 400 MeV the observations can be described in the synchrotron forward-shock scenario,  between $\sim$ 0.6 - 1 GeV the contribution of  both processes would  be relevant,  and  beyond the synchrotron limit,  the observations would be entirely explained by SSC process for the set of values used. \\
The bottom right-hand panel of Fig. \ref{fig1:wind} shows that the maximum SSC flux  lies at $\sim$ 100 GeV,  making it possible to detect the VHE emission in observatories where the sensitivity is maximum  at hundreds of GeV but not  in those observatories where the maximum sensitivity lies in few TeVs \citep[e.g. HAWC;][]{2012APh....35..641A}. Similar to the case of the stratified medium, the SSC flux at 1 TeV decreases between two and three orders of magnitude in comparison with the flux at 100 GeV.\\
Given the minimum and cooling electron Lorentz factors (eq. \ref{elec_factor}), the synchrotron and SSC luminosity ratio can be computed as \citep{2001ApJ...548..787S}
\be
\frac{L^{\rm ssc}_\nu}{L^{\rm syn}_\nu}\simeq 5.1\times 10^{-2}\, n_{-1}\, R_{\rm dec, 17}\,\gamma_{\rm c,4}^2\left(\frac{\gamma_{\rm c,4}}{\gamma_{\rm m,2}}\right)^{1-p}\,,
\ee
where $R_{\rm dec}=4.22\times10^{17}\,{\rm cm} \left(\frac{1+z}{1.5} \right)^{-\frac14}\, n^{-\frac14}_{-1}\,E^{\frac14}_{53.5}\,t^{\frac14}_3$ is the deceleration radius.   Again,  it is worth mentioning that the synchrotron and SSC luminosity ratio depends on $Y$ through $\gamma_{\rm c}$.   In the case of the uniform medium, only 5\% of the synchrotron luminosity is up-scattered by SSC emission.
\subsection{The stratified-to-homogeneous afterglow transition}
\begin{figure}[h!]
{ \centering
\resizebox*{0.48\textwidth}{0.3\textheight}
{\includegraphics{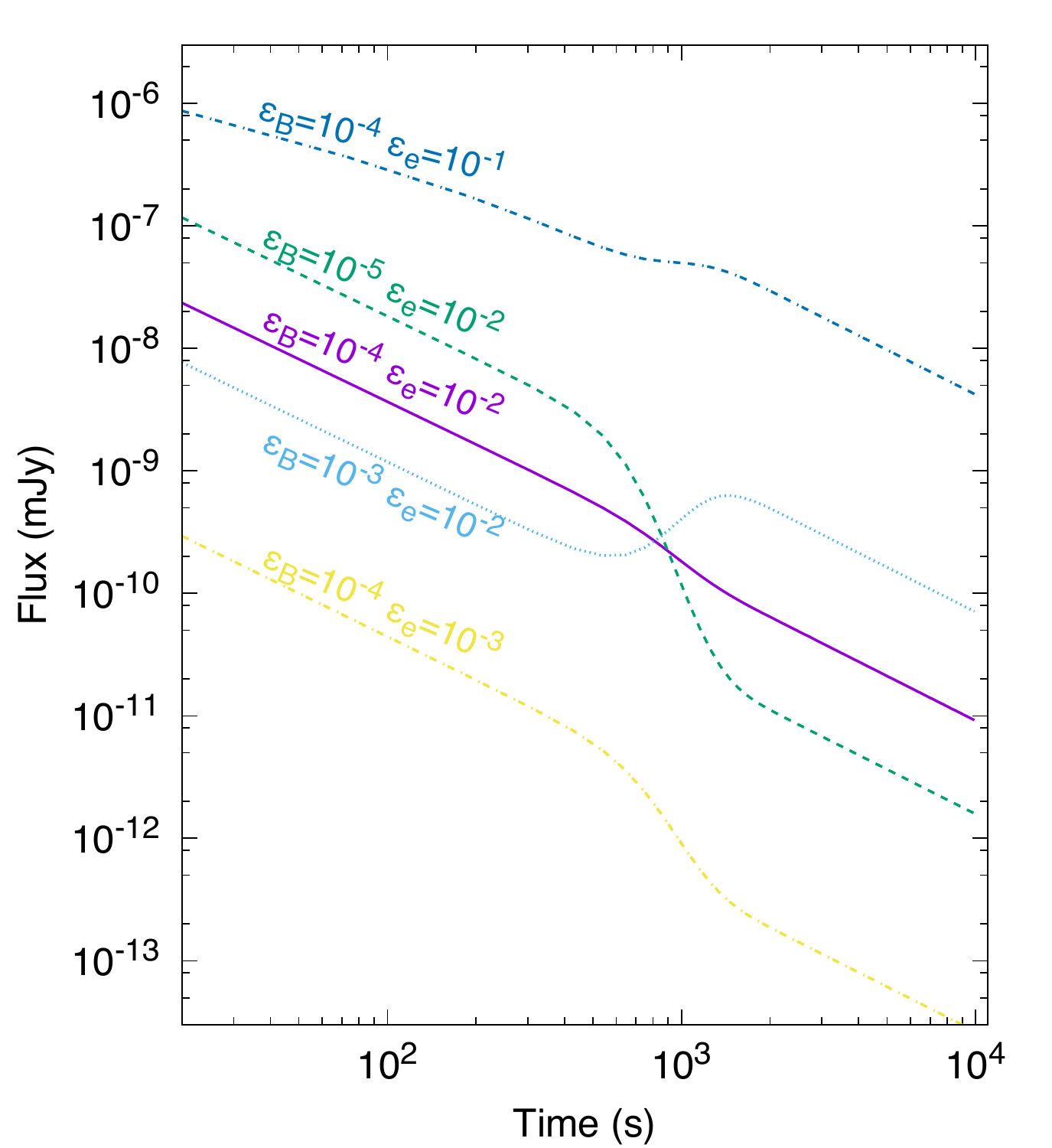}}
\resizebox*{0.48\textwidth}{0.3\textheight}
{\includegraphics{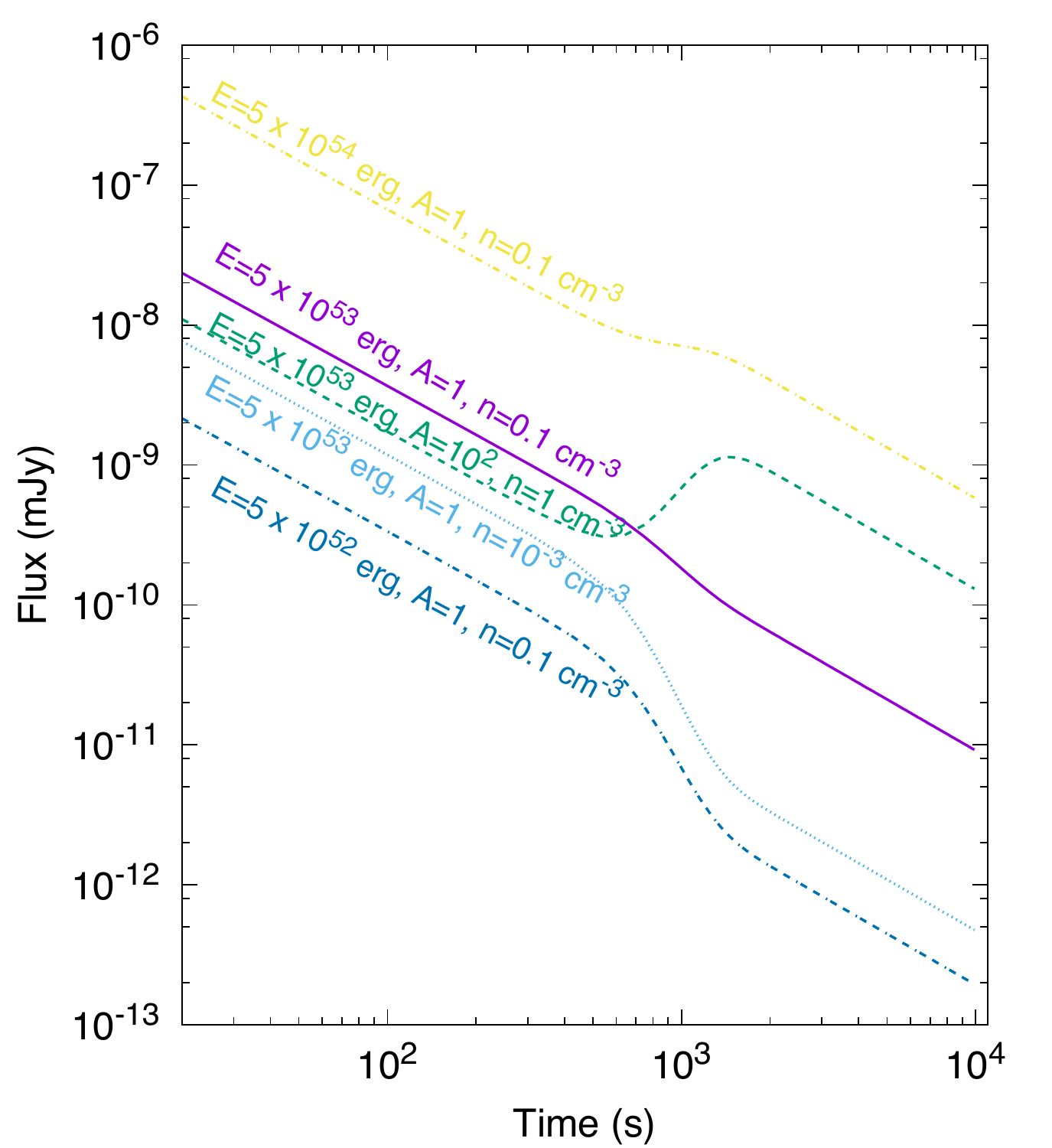}}
\resizebox*{0.48\textwidth}{0.3\textheight}
{\includegraphics{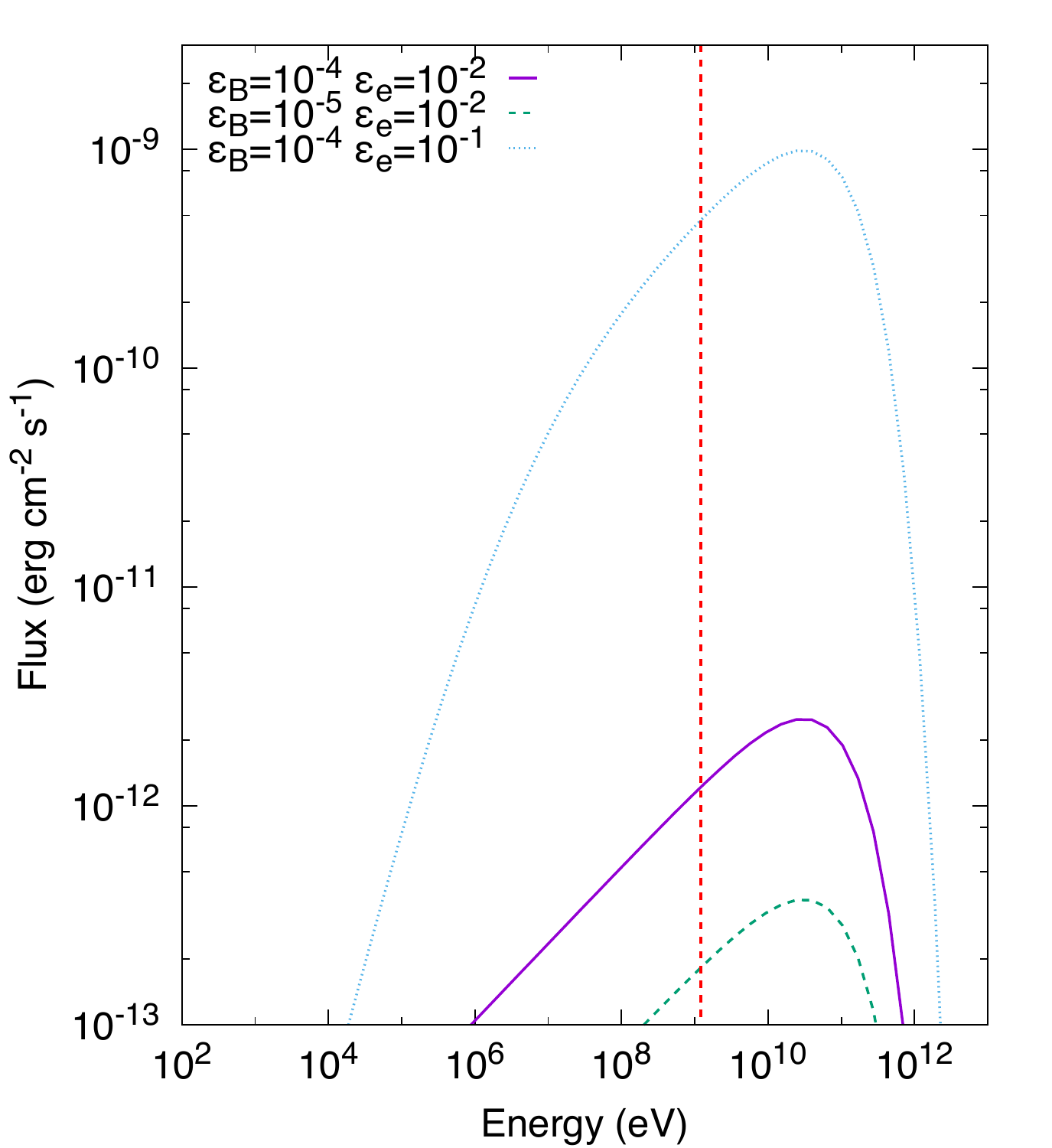}}
\resizebox*{0.48\textwidth}{0.3\textheight}
{\includegraphics{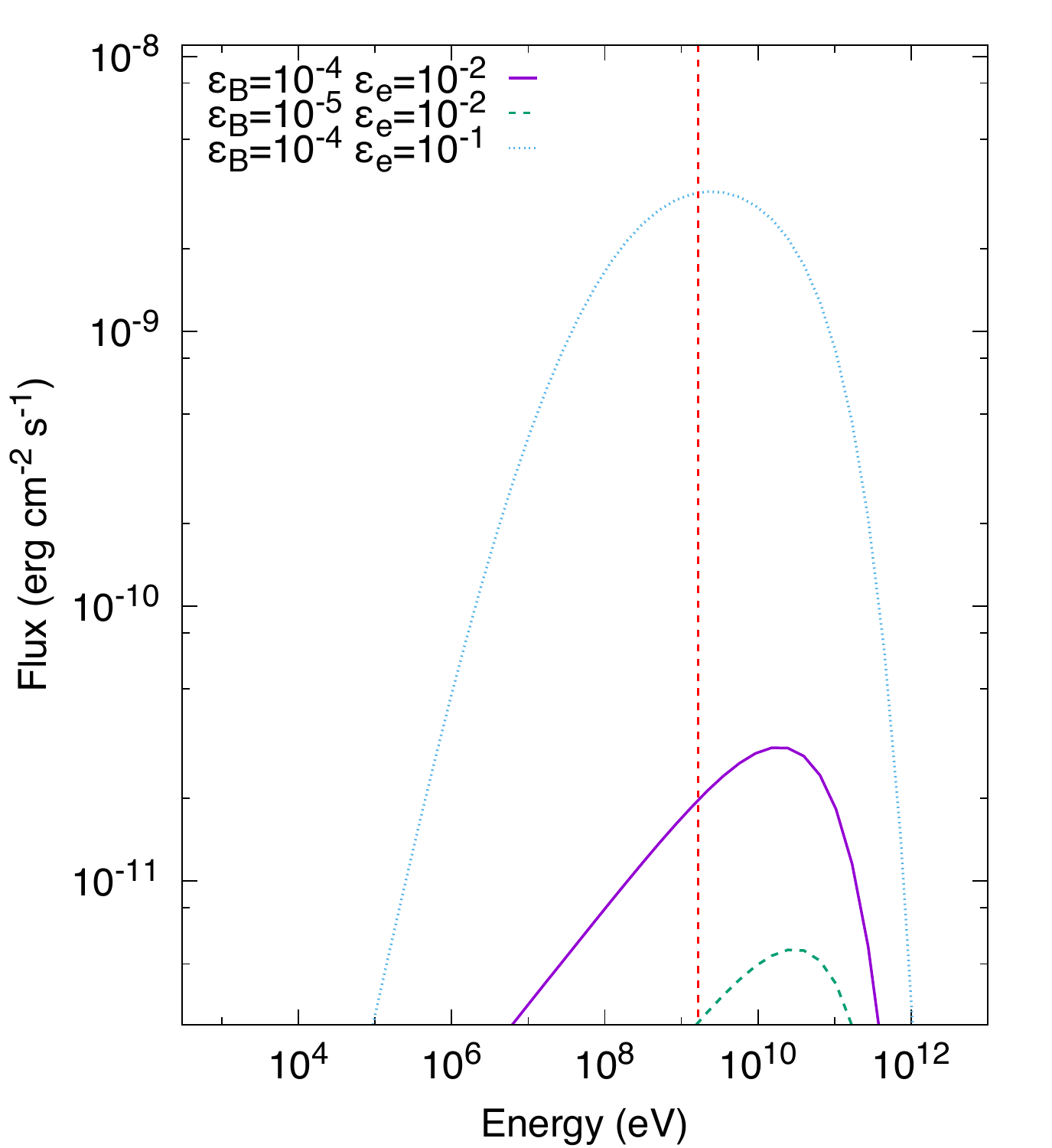}}
}
\caption{ Figure shows the SSC light curves and spectra during the afterglow transition between the stratified and homogeneous medium (considered here at 1000 s) for relevant values of GRB afterglows.   In the top left-hand panel, light curves are obtained for $E= 5\times10^{53}\,{\rm erg}$, $A_{\rm w} = 1$ and $n= 1\,{\rm cm^{-3}}$, and in the  top right-hand panel these are for $\varepsilon_{\rm B}=10^{-4}$ and $\varepsilon_{\rm B}=10^{-2}$.  The  Top panels show the SSC light curves for $\epsilon_\gamma=100\,{\rm GeV}$ and the bottom ones the SSC spectra computed in the stratified (left) and homogeneous (right) medium. The SSC spectra in the stratified medium are computed for  $t=900\,{\rm s}$ and in the homogeneous medium  for $t=1100\,{\rm s}$.   The red dashed line corresponds to the maximum energy radiated by synchrotron.}
 \label{fig2:transition}
\end{figure}

\begin{figure}[h!]
{ \centering
\resizebox*{1.\textwidth}{0.5\textheight}
{\includegraphics{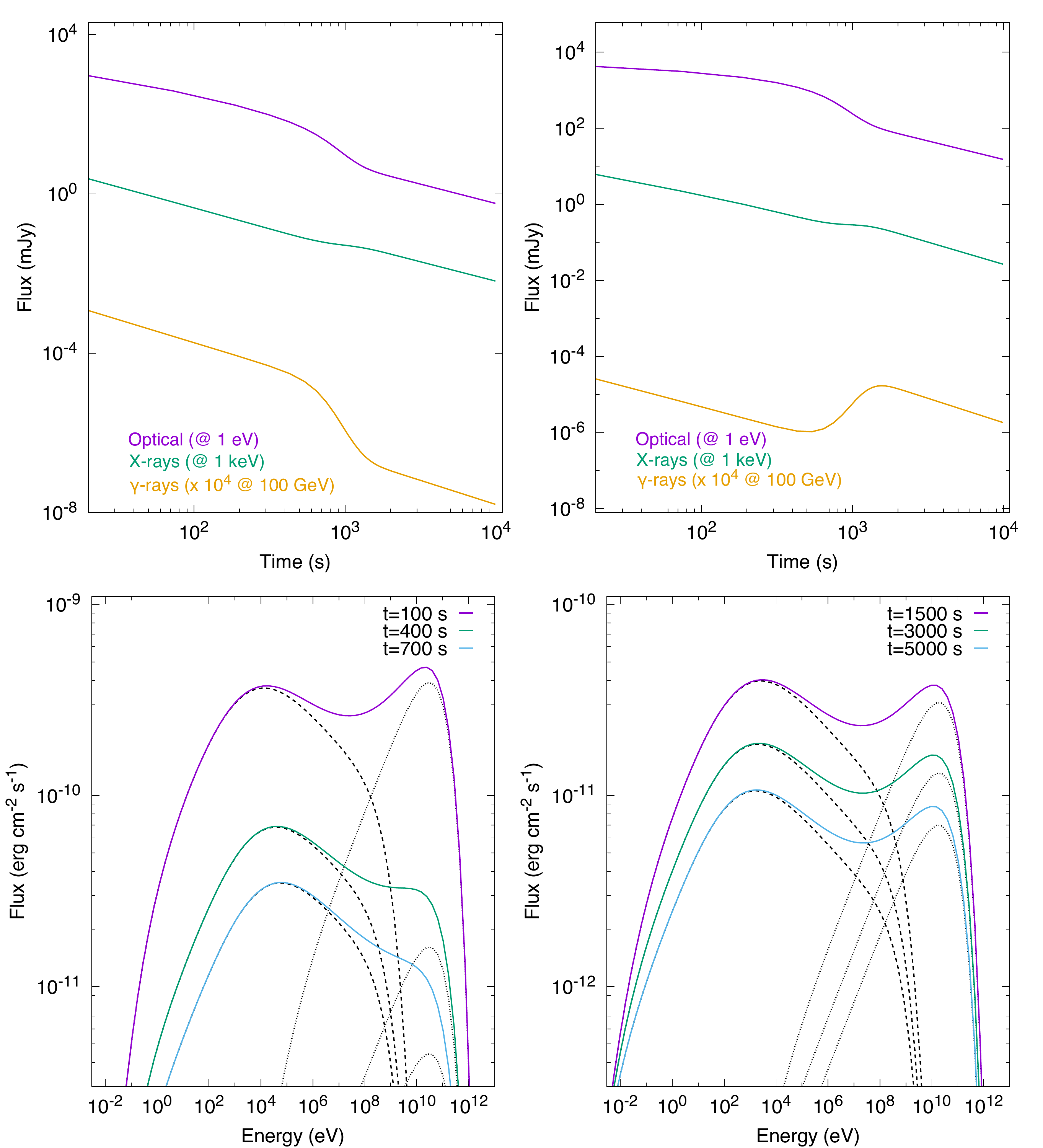}}
}
\caption{Figure shows the SSC and synchrotron light curves (SSC model) and the broadband SEDs during the afterglow transition between the stratified and homogeneous medium. The top panels show the synchrotron light curves of optical and X-ray bands at 1 eV and 1 keV, and the SSC light curves  of $\gamma$-rays at  100 GeV.   In the top left-hand panel, the light curves are computed for $E= 5\times10^{53}\,{\rm erg}$, $A_{\rm w} = 0.1$, $n= 0.1\,{\rm cm^{-3}}$, $\varepsilon_{\rm B}= 10^{-4}$,  $\varepsilon_{\rm e}= 10^{-2}$  and $\xi=1$, and in the top right-hand panel, these are obtained for $E= 10^{54}\,{\rm erg}$, $A_{\rm w} = 1$, $n= 1\,{\rm cm^{-3}}$, $\varepsilon_{\rm B}= 10^{-3}$,  $\varepsilon_{\rm e}= 10^{-2}$ and $\xi=0.5$. The bottom panels show the broadband SEDs (both synchrotron - dashed, and SSC - dotted) computed in the stratified medium for t= 100, 400 and 700 s (left) and the homogeneous  medium for t= 1500, 3000 and 5000 s (right). The stratified-to-homogeneous transition is considered at 1000 s, as in Fig. \ref{fig2:transition}.  In the bottom panels,  the SEDs are computed for $E= 5\times10^{53}\,{\rm erg}$,  $\varepsilon_{\rm B}= 10^{-4}$ and $\varepsilon_{\rm e}= 10^{-2}$, with the densities of $A_{\rm W}=0.1$ (left) and $n= 1\,{\rm cm^{-3}}$ (right).}
 \label{fig3:syn_ssc}
\end{figure}

Figure \ref{fig2:transition} shows the SSC light curves and spectra during the afterglow transition between the stratified and homogeneous medium for typical values in the ranges: $5\times10^{52} \leq E\leq 5\times10^{54}\,{\rm erg}$, $0.1\,{\rm cm^{-3}} \leq n\leq 1\,{\rm cm^{-3}}$, $10^{-2}\leq A_{\rm w} \leq 1$,  $10^{-5}\leq \varepsilon_{\rm B}\leq 10^{-3}$ and  $10^{-3}\leq \varepsilon_{\rm e}\leq 10^{-1}$.  The  top panels show the SSC light curves for $\epsilon_\gamma=100\,{\rm GeV}$.  The stratified-to-homogeneous transition radius can be written as \citep[e.g., see][]{2017ApJ...848...15F}
\be
R_{\rm tr}\simeq 5.1\times 10^{18}\,{\rm cm}\, \dot{M}^{\frac{3}{10}}_{-6}\,v^{\frac{1}{10}}_{W,8}\,n^{-\frac{3}{10}}\,t^{\frac25}_{\star,5}\,,
\ee
where $t_\star$ is the lifetime of the star phase for $n=1\,{\rm cm^{-3}}$.   In our analysis, we have considered the stratified-to-homogeneous afterglow transition at 1000 s which correspond to a deceleration radius of $\simeq 4.2\times10^{18}\,{\rm cm}$ for $E=5\times 10^{54}\,{\rm erg}$ and $A_{\rm W}=10^{-1}$ or $E=5\times 10^{53}\,{\rm erg}$ and $A_{\rm W}=10^{-2}$.  In the top left-hand panel, the light curves are computed for  $E= 5\times10^{53}\,{\rm erg}$, $A_{\rm w} = 1$ and $n= 1\,{\rm cm^{-3}}$  and in the top right-hand panel the light curves are obtained for $\varepsilon_{\rm B}= 10^{-4}$ and  $\varepsilon_{\rm e}= 10^{-2}$.  These panels show that depending on the parameter values, the afterglow transition can be quite noticeable.    For instance, in the purple curve there is actually a smoother transition (which is harder to detect) compared to some of the others.   The top left-hand panel shows that SSC fluxes increase as $\varepsilon_{\rm e}$ increases in the stratified and the homogeneous medium;  higher values of $\varepsilon_{\rm e}$ make SSC emission more favorable to be detected \citep[indeed such values are expected to be common in GRB afterglows, see e.g.,][]{2014ApJ...785...29S, 2017MNRAS.472.3161B}.  Moreover, the SSC fluxes  increase as  $\varepsilon_{\rm B}$ decreases in the stratified but not in the homogeneous medium. The top right-hand panel shows that the SSC fluxes increase as $E$, $n$ and $A_{\rm W}$  increase in both the stratified and homogeneous medium.\\ 
The  bottom panels of Fig. \ref{fig2:transition} show the SSC spectra computed in the stratified medium for $t=900\,{\rm s}$ (left  panel) and in the homogeneous medium for $t=1100\,{\rm s}$ (right panel).  The red dashed line represents the synchrotron limit.  The bottom panels show that these spectra increase dramatically as $\epsilon_{\rm e}$ increases and slightly as $\epsilon_{\rm B}$ increases. The SSC light curves with the same colors (parameter values) represent the evolution from the stratified to homogeneous medium. As a consequence of this transition, one can observe that SSC fluxes increase up to more than one order of magnitude.\\
\\
Figure \ref{fig3:syn_ssc} shows the synchrotron and SSC light curves, and the SEDs during the afterglow transition between the stratified and homogeneous medium. The spectrum and  light curves of synchrotron emission have been included with the purpose of performing a multi-wavelength analysis.
\vspace{0.5cm}
\subsubsection{Multi-wavelength Light Curves Analysis}
The synchrotron light curves of optical and X-ray bands at 1 eV and 1 keV, and the SSC light curves  of $\gamma$-rays at  100 GeV  are shown in the top panels of Figure \ref{fig3:syn_ssc}.   In both panels  it can be seen that while optical and X-ray fluxes display the same behavior, $\gamma$-rays exhibit  a different one.   For the given parameter values,  during the afterglow transition the $\gamma$-ray flux can decrease (left  panel) or increase (right panel).    Taking into consideration the parameter values used in the top left-hand panel, for the stratified medium,  the synchrotron and the SSC spectral breaks are  $\epsilon^{\rm syn}_{\rm m}=5.1\times 10^{-2}\,{\rm eV}$, $\epsilon^{\rm syn}_{\rm c}=2.3\,{\rm keV}$,  $\epsilon^{\rm ssc}_{\rm m}=15.1\,{\rm keV}$, $\epsilon^{\rm ssc}_{\rm c}=97.7\,{\rm TeV}$, $\epsilon^{\rm KN}_{\rm c}=8.8\,{\rm TeV}$  and  for the homogeneous medium, these breaks are $\epsilon^{\rm syn}_{\rm m}=2.2\times 10^{-2}\,{\rm eV}$, $\epsilon^{\rm syn}_{\rm c}=1.8\,{\rm keV}$,  $\epsilon^{\rm ssc}_{\rm m}=3.6\,{\rm keV}$, $\epsilon^{\rm ssc}_{\rm c}=8.3\,{\rm TeV}$, $\epsilon^{\rm KN}_{\rm c}=2.4\,{\rm TeV}$.  Considering  the parameter values used in the right-hand panel,   the SSC and synchrotron spectral breaks computed in the stratified medium are $\epsilon^{\rm syn}_{\rm m}=4.3\,{\rm eV}$, $\epsilon^{\rm syn}_{\rm c}=3.5\times 10^{-2}\,{\rm eV}$,  $\epsilon^{\rm ssc}_{\rm m}=0.5\,{\rm MeV}$, $\epsilon^{\rm ssc}_{\rm c}=0.5\,{\rm keV}$, $\epsilon^{\rm KN}_{\rm c}=4.9\,{\rm GeV}$ and these breaks computed in the homogeneous medium are $\epsilon^{\rm syn}_{\rm m}=8.5\times 10^{-2}\,{\rm eV}$, $\epsilon^{\rm syn}_{\rm c}=14.3\,{\rm eV}$,  $\epsilon^{\rm ssc}_{\rm m}=10.8\,{\rm keV}$, $\epsilon^{\rm ssc}_{\rm c}=0.2\,{\rm GeV}$, $\epsilon^{\rm KN}_{\rm c}=2.6\,{\rm GeV}$.   Therefore,  from stratified-to-homogeneous medium the optical flux evolves in the second PL segment and  the X-rays  in the third PL segment  of synchrotron model.  During this transition phase the temporal index of the third PL segment of synchrotron emission ($\propto t^{-\frac{3p-2}{4}}$) does not vary, and the second PL segment  varies from $\propto t^{-\frac{3p-1}{4}}$ to $\propto t^{-\frac{3p-3}{4}}$. However, an alternative interpretation different to the afterglow transition could be given in terms of the reverse-shock emission.  In this framework, the X-rays are not altered and the optical flash is detected with a decay flux of $\propto t^{-\frac{73p+21}{96}}$ and $\propto t^{-\frac{27p+7}{35}}$ for the thick and thin shell, respectively \citep{2000ApJ...545..807K}.   In this case, the analysis of the SSC light curve would be very useful in order to differentiate both interpretations. With the parameters given,  the $\gamma$-ray evolves in the second PL segment close to the afterglow transition, from $\propto t^{-p}$ to $\propto t^{-\frac18}$.      With the given parameters, the gamma-rays evolves in the second PL segment as the medium changes from wind to ISM from $\propto t^{-p}$ or $\propto t^{-\frac18}$, respectively; if the medium does not have this transition, then the flux would not show this particular break in the light curve.   It is worth noting that while the afterglow transition is imperceptible for the synchrotron light curve at 1 keV,  it presents a discontinuity quite evident for the SSC light curve at 100 GeV.\\
%
%
%
%
\subsubsection{The broadband Spectral Energy Distribution Analysis}
%
The bottom panels of Fig. \ref{fig3:syn_ssc}  show the SEDs computed in the stratified (left panel) and homogeneous (right  panel) medium for a transition at 1000 s.  For the case of stratified medium, we assume each SED at 100, 400 and 700 s and for the case of homogeneous medium, we calculate each SED at 1500, 3000 and 5000 s. Densities with values of  $A=0.1$ and $n=1\,{\rm cm^{-3}}$ are chosen for the stratified and homogeneous medium, respectively, and in both cases we use the same values $E= 5\times10^{53}\,{\rm erg}$, $\varepsilon_{\rm B}= 10^{-4}$ and  $\varepsilon_{\rm e}= 10^{-2}$  and $\xi=1.0$.  The principal features are: i) While the ratio between the maximum synchrotron and the SSC fluxes decreases drastically in the stratified medium, it remains quasi-constant in the homogeneous medium.   ii)  While the synchrotron peak is shifted to higher energies as time increases in the stratified medium, it evolves quite slowly with time in the homogeneous medium. iii) While the maximum value of the SSC flux decreases  quickly with time for the stratified medium, this value decreases gradually for the homogeneous medium.   iv) An increase in the synchrotron and SSC fluxes is  seen during the afterglow transition.  At $t=700\,{\rm s}$ the maximum values of synchrotron and SSC fluxes are  $3.5\times 10^{-11}$ and $4.2\times 10^{-12}\,{\rm erg\,cm^{-2}}\,s^{-1}$, respectively, and at $t=1500\,{\rm s}$ the maximum values of synchrotron and SSC fluxes are $4.2\times 10^{-11}$ and $3.1\times 10^{-11}\,{\rm erg\,cm^{-2}}\,s^{-1}$ and  iv) The evolution of the SED structures in both stratified and homogeneous medium  are different.  These characteristics could help identify if the transition phase exists or  it is simply  associated to a distinct scenario.
%
%
%
\section{Application to GRB 190114C}
\subsection{Multi-wavelength Observations and previous analysis} \label{sec:observations}
GRB 190114C was triggered by the Burst Area Telescope (BAT) instrument onboard the Swift satellite on January 14, 2019 at 20:57:06.012 UTC \citep{2019GCN.23688....1G}.   VHE photons with energies above 300 GeV were detected from this burst with a significance of 20$\sigma$ by the MAGIC telescope for more than 20 minutes \citep{2019GCN.23701....1M}.  GRB 190114C was followed up by a massive observational campaign with instruments onboard satellites and ground telescopes covering a large fraction of the electromagnetic spectrum  \citep[see][and references therein]{2019ApJ...879L..26F}.  The host galaxy of GRB 190114C was located and confirmed to have a redshift of  $z$ = 0.42 \citep{2019GCN.23692....1U,2019GCN.23695....1S}.\\
Recently, \cite{2019ApJ...879L..26F} showed that the LAT light curve of GRB 190114C exhibited similar features to other bright LAT-detected bursts. Together with  the multi-wavelength observations, the long-lived LAT, GBM, X-ray, optical and radio emissions were consistent with the standard synchrotron forward-shock model that evolves from a stratified to a homogeneous medium with an afterglow transition at $\sim 400$ s.  These authors showed that the high-energy photons were produced in the deceleration phase of the relativistic outflow and also that some additional processes to synchrotron in the forward shocks  should be considered to properly describe the LAT photons with energies beyond the synchrotron limit.  Here, we use the SSC process to interpret the photons beyond this synchrotron limit.\\
\subsection{Estimation of SSC light curves and VHE photons beyond the synchrotron limit}
\begin{figure}[h!]
{ \centering
\resizebox*{0.5\textwidth}{0.4\textheight}
{\includegraphics{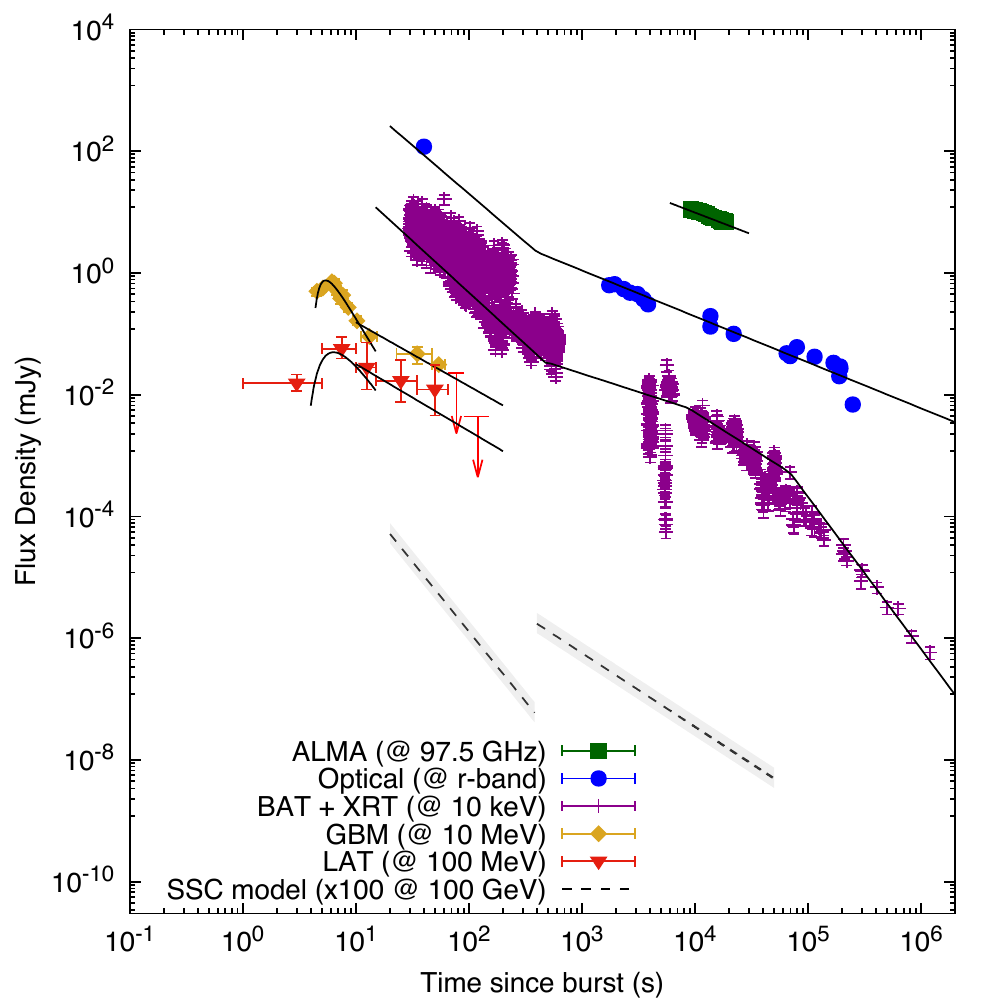}}
\resizebox*{0.5\textwidth}{0.4\textheight}
{\includegraphics{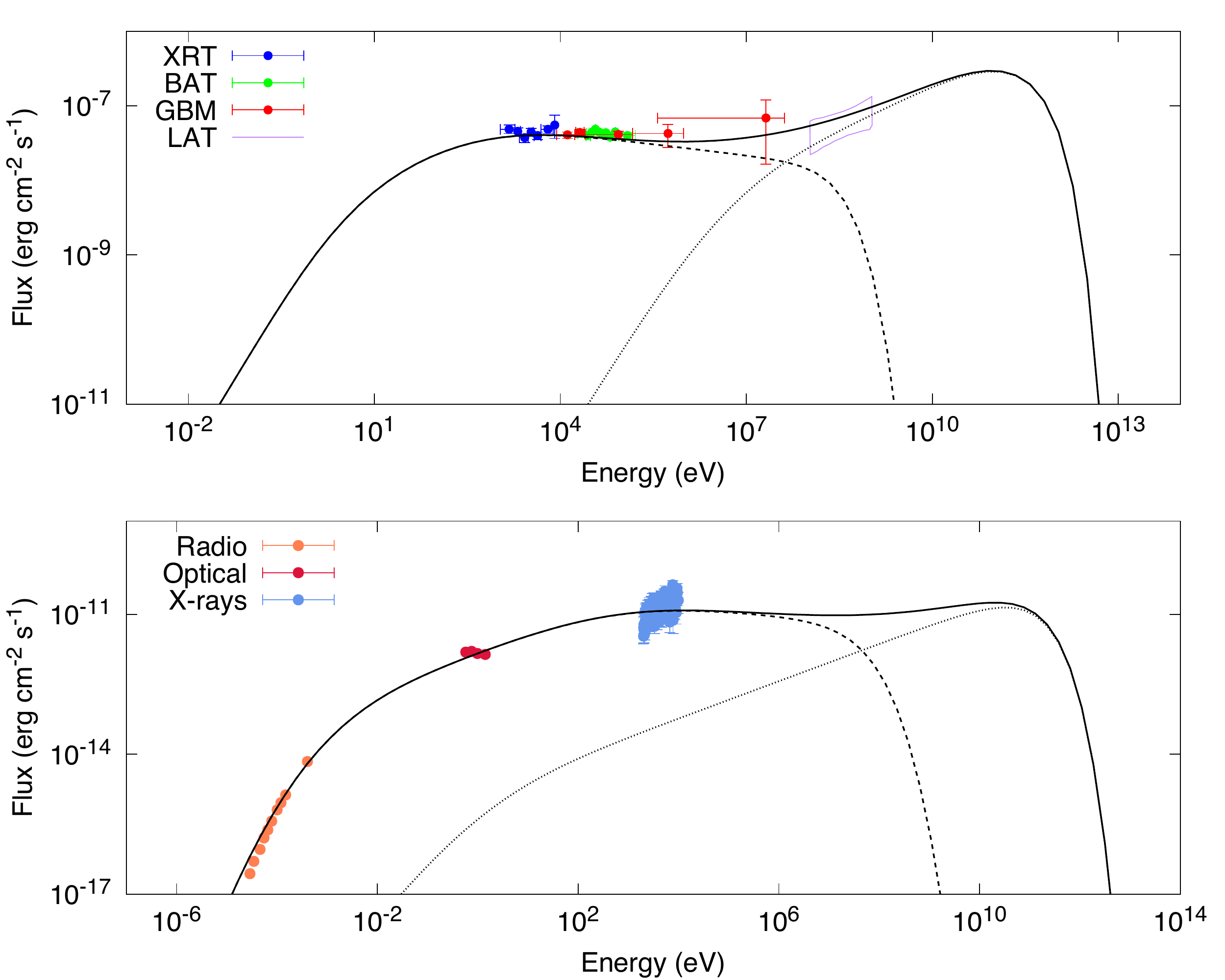}}
}%
\caption{The SSC light curves (left) and SEDs (right) of GRB 190114C are shown. The synchrotron and SSC curves in a stratified and homogeneous medium were derived using the best-fit parameters reported in  \cite{2019ApJ...879L..26F}.  The broadband SEDs are built from two time intervals 66 - 92 s (above) and 0.2 days (below).  The left-hand panel is adapted from \cite{2019ApJ...879L..26F}.  The effect of the extragalactic background light (EBL) absorption proposed by \cite{2017A&A...603A..34F} is taken into account.  Radio, optical, X-ray, GBM and LAT data were taken from \cite{2019arXiv190407261L,  2019ApJ...879L..26F, 2019arXiv190201861R}.  The values used to obtain the SSC light curves and SEDs 
are  $A_{\rm W}=6\times 10^{-2}$, $n=1\,{\rm cm^{-3}}$, $E=2\times 10^{54}\,{\rm erg}$,  $\varepsilon_{\rm B}=5\times10^{-6}$,  $\varepsilon_{\rm e}=10^{-2}$ and $p=2.3$ \citep{2019ApJ...879L..26F}.}
 \label{fig2:GRB190114C}
\end{figure}

Using the best-fit values reported in \cite{2019ApJ...879L..26F}, the SSC light curves were calculated.   The left-hand panel in Figure \ref{fig2:GRB190114C} shows the SSC light curves at 100 GeV  in a stratified and homogeneous medium. This panel was adapted  from  \cite{2019ApJ...879L..26F}. The effect of the EBL absorption proposed by \cite{2017A&A...603A..34F}  was included.\\
The values of transition times between fast- and slow-cooling regime are 0.2 and 0.09 s for the stratified and homogenous medium, respectively.  The values of the characteristic and cutoff SSC breaks calculated in the stratified medium are $\epsilon^{\rm ssc}_{\rm m} \simeq0.2\,{\rm MeV}$ and  $\epsilon^{\rm ssc}_{c}\simeq4.2\,\times 10^4\,{\rm keV}$ at 100 s, and  in the homogeneous medium are $\epsilon^{\rm ssc}_{\rm m} \simeq1.6\,{\rm keV}$ and  $\epsilon^{\rm ssc}_{\rm c}\simeq1.6\,{\rm keV}$, respectively,  at 1000 s. Therefore, in both cases the SSC light curves evolves in the second PL segment of the slow-cooling regime.  The break energies in the KN regime are 200.7 TeV and 868.1 GeV at 100 and $10^3\,{\rm s}$, respectively.  The highest energy photons reported by LAT and MAGIC collaboration are below the KN regime which agrees with the description of the SSC light curves.  We emphasize that the parameters obtained with the MCMC code from the broadband modeling of the multi-wavelength observations may be changed somewhat when the KN effects are included, but the SSC emission itself will not be strongly affected.  Therefore, VHE photons beyond the synchrotron limit can be explained through the second PL segment of the SSC emission in the slow-cooling regime.\\
In our model, the SSC emission decays steeper in the stratified than the homogeneous medium. However, during the stratified-to-homogeneous transition the SSC flux suddenly increases by $\sim$ one order of magnitude.  This allows that the SSC component could be detected during a longer time.\\
\\
\cite{2012APh....35..641A} presented the HAWC sensitivity of the scaler system to GRBs for several declinations and energies (at which this observatory is sensitive).    At 1 TeV and for a  power-law index of $p=2.15\pm 0.35$,  the HAWC sensitivities for the declinations of $1.0\geq \cos{\theta}>0.9$, $0.9\geq \cos{\theta}>0.8$, $0.8\geq \cos{\theta}>0.7$ and $0.7\geq \cos{\theta}>0.6$ are  $\sim(2.2\times 10^{-7} -  1.1\times 10^{-6})$ mJy,  $\sim(0.4\times 10^{-6} -  2.5\times 10^{-6})$ mJy,  $\sim(0.8\times 10^{-6} -  0.7\times 10^{-5})$ mJy and  $\sim(0.8\times 10^{-5} -  0.7\times 10^{-4})$ mJy, respectively. Taking into account the attenuation factor $\sim10^{-2}$ due to EBL at 1 TeV, the SSC flux would be $\sim 10^{-10}$ mJy at $10^2\, {\rm s}$ and $\sim 10^{-12}$ mJy at $10^3\, {\rm s}$ for the stratified and homogeneous medium, respectively.  These values are well below the HAWC sensitivity for any declination.  
It shows that, with our model, GRB 190114C could not be detected by the HAWC observatory, even if this burst would have been located at the HAWC's field of view.\\
\\
\cite{2013APh....43..348F}  and \cite{2016CRPhy..17..617P} presented and discussed the sensitivity to transient sources as a function of duration of High Energy Stereoscopic System (HESS) CT5 and Cherenkov Telescope Array (CTA) telescopes for distinct energy thresholds. At 500 s, the HESS CT5 and CTA sensitivities for energy thresholds of  75 and 80 GeVs are $\sim 2 \times 10^{-10}$ mJy and $\sim7\times 10^{-10}$ mJy, respectively. Therefore,  if  GRB 190114C would have been fast located by HESS CT5 telescope, this burst would have been detected by HESS in accordance with our model. Similarly, bursts with similar features of GRB 190114C are perfect candidates for detection with future VHE facilities \citep[e.g. CTA;][]{2013APh....43..348F}. \\ 
\subsection{The broadband Spectral Energy Distributions}
The right-hand panels of Fig.~\ref{fig2:GRB190114C} show the  SEDs at 66 - 92 s  (top panel) and 0.2 days (bottom panel).   The synchrotron and SSC curves in a stratified (above) and homogeneous (below) medium were derived using the best-fit parameters reported in  \cite{2019ApJ...879L..26F}.  The EBL model introduced in \cite{2017A&A...603A..34F} was used.  Radio, optical, X-ray, GBM and LAT data were taken from \cite{2019arXiv190407261L,  2019ApJ...879L..26F, 2019arXiv190201861R}.\\
The  top panel shows that synchrotron emission describes the  optical to LAT energy range and SSC emission contributes significantly to the LAT observations.  The bottom panel shows  that synchrotron emission explains the ratio to X-ray data points. The flux ratio at the peaks are $F^{\rm syn}_\nu/F^{\rm ssc}_\nu\simeq10$ and $\simeq 1$ at   66 - 90 s and 0.2 days, respectively. The SEDs  can be explained through the evolution of synchrotron and SSC emissions in the stratified and the homogeneous medium. The decay of SSC emission is steeper than synchrotron radiation;  in the stratified medium the decay of SSC and synchrotron evolve as $\propto t^{-2.30}$ and $\propto t^{-1.47}$, respectively,  and in  the homogeneous medium as $\propto t^{-1.21}$ and $t^{-0.97}$, respectively. \\
\subsection{Why GRB 190114C is special in comparison to other LAT-detected bursts}
%
%
The VHE flux above $\sim$ 100 GeV begins to be attenuated by pair production with EBL photons \citep{1966PhRvL..16..252G}. The SSC flux observed is attenuated by $\exp[-\tau_{\gamma\gamma} (z)]$ with $\tau_{\gamma\gamma}(z)$ the photon-photon opacity as a function of redshift.  Using the values of the opacities reported in \cite{2017A&A...603A..34F}, VHE emission with photons at 300 GeV (1 TeV)  is attenuated by $0.15$ ($2.1\times 10^{-3}$) and $6.6\times 10^{-3}$ ($2.2\times 10^{-6}$)  for  z=0.5 and 1, respectively. In the particular case of GRB 190114C, a low-redshift of  z=0.42 allowed the detection by an  Imaging Atmospheric Cherenkov Telescope such as MAGIC.\\

We show that changes in density of the circumburst medium leads to an increase in the SSC emission. The afterglow transition reported in GRB 190114C  allowed for enhanced VHE emission increased and hence was detected for a longer period.\\ 

%
The set of the best-fit parameters as reported for GRB 190114C made  more favorable its detection by the MAGIC telescopes. As follows, we enumerate each one:
%

\begin{enumerate}
\item The SSC emission  peaked below the KN regime, otherwise it would be drastically attenuated. 

\item  The peak of the SSC emission was reproduced at hundreds of GeVs, where MAGIC is more sensitive and the attenuation by EBL is small. Other  configurations of parameters lead to peaks at few TeVs where the EBL absorption is much higher and therefore more difficult to detect by by Imaging Atmospheric Cherenkov Telescopes (IACTs). \\

\item With the parameters given, the SSC flux evolved in the second PL segment of slow-cooling regime. In this case, we show that the SSC flux increases as the densities in both the stratified and homogeneous media increase. The values of densities in both cases  make the detection of VHE flux more favorable.

\item   The LAT light curve of GRB 190114C exhibited  similar characteristics to other powerful bursts detected by Fermi LAT \citep[see][]{2019ApJ...879L..26F}.\footnote{GRB 080916C, GRB 090510, GRB 090902B, GRB 090926A, GRB 110721A, GRB 110731A, GRB 130427A and GRB 160625B and others \citep{2009Sci...323.1688A,  2010ApJ...716.1178A, 2009ApJ...706L.138A, 2011ApJ...729..114A, 2013ApJ...763...71A,2017ApJ...848...94F, 2013ApJ...763...71A, 2014Sci...343...42A, 2017ApJ...848...15F}} These authors showed that GRB 190114C corresponded to one of the more powerful bursts during the first hundreds of seconds (early afterglow). In this work  we show that higher values of the equivalent kinetic energy make the SSC emission more favorable to be detected. Therefore, the total energy reported of this burst favored to its detection.

\end{enumerate}

Notwithstanding attempts to detect the VHE emission at hundreds of GeVs  by IACTs have been an arduous task because the time needed to locate the burst is longer than the duration of the prompt  and early-afterglow emission, only one detection has been reported   \citep[GRB 190114C;][]{2019GCN.23701....1M}. During the last two decades,  only upper VHE limits  have been derived by these telescopes \citep[e.g. see,][]{Albert_MAGIC, Aleksi_GRB090102,  Aharonian_2009HESS, Aharonian_GRB060602B, HESS_GRB100621A, Acciari_VERITAS,  Bartoli_ARGO, GRB150323A_Abeysekara}.    We conclude that the conditions to locate promptly the early afterglow of GRB 190114C by the MAGIC telescope together with the low-redshift and favorable set of parameters made its detection possible. We want to highlight that no other LAT-detected burst complies with all the requirements mentioned above.  It is worth noting that although GRB 130427A was closer and more energetic than GRB 190114C \citep{2014Sci...343...42A}, it was not located rapidly enough to catch the early afterglow by IACTs.\footnote{VERITAS started follow-up observations of GRB 130427A $\sim$ 20 hours after the trigger time \citep{2014ApJ...795L...3A}.}

\section{Conclusions}
We have computed the SSC light curves for a stratified stellar-wind that transitions to an homogeneous ISM-like medium, taking into account the synchrotron forward-shock models introduced in \cite{1998ApJ...497L..17S},  \cite{2000ApJ...536..195C}  and \cite{2000ApJ...543...66P}.   The break energy in the KN regime was obtained.   The attenuation produced by the EBL absorption is introduced in accordance with the model presented in \cite{2017A&A...603A..34F}.  The intrinsic attenuation by $e^\pm$ pair production (opacity) is not taken into account  because during the deceleration phase  it is much less than unity ($\tau_{\rm \gamma\gamma,in}\ll 1$).\\ 
In general, we compute the SSC light curves for a stratified and homogeneous medium at 10 and 100 GeV and compare them with the LAT and MAGIC sensitivities.  We show that depending on the parameter values, the SSC light curves are above the LAT and MAGIC sensitivities.  We calculate the SSC light curves during the afterglow transition and show that for this transition to be well-identified, it is necessary not only to observe the synchrotron, but also the SSC emission. For instance, the SSC emission can help us to discriminate between the stratified-to-homogeneous afterglow transition and a reverse-shock scenario.   We have computed the SED in the stratified and homogeneous medium and also discussed their differences.\\
We emphasize that the equations of synchrotron flux are degenerate in parameters such that for an entirely distinct set of parameters same results can be obtained. Therefore, this result is not unique, but it is a possible solution for GRB 190114C.  It is worth noting that  if the increase of the observed SSC flux around $\sim$ 400 s is not exhibited, then other set of parameters to describe  GRB 190114C is required or an alternative scenario would have to be evoked. \\
Using the best-fit parameters reported for GRB 190114C, we have estimated the SSC light curves and fitted the SEDs  for two epochs  66 - 92 s  and 0.2 days.  We show that  SSC process could explain the VHE  photons beyond the synchrotron limit in GRB 190114C.\\
\\
Recently, \cite{2019arXiv190511312W} described the broadband SED of GRB 190114C with a SSC model for a homogeneous medium using the optical, X-ray and LAT data between 50 - 150 s. They concluded that the detection of sub-TeV photons is attributed to the large burst energy and low redshift.  \cite{2019arXiv190508285D} studied the physical conditions of the afterglow required for explaining the sub-TeV photons in GRB 190114C. These authors found that the Comptonization of X-ray photons at the border between Thompson and KN regime  with a bulk and electron  Lorentz factor of $\simeq 100$ and $\gamma_c\simeq 10^4$ could described the MAGIC detection. In our current work, in addition to study the evolution of SSC light curves during the stratified-to-homogeneous afterglow transition as reported in \cite{2019ApJ...879L..26F}, we have  interpreted the  photons beyond the synchrotron limit in the SSC framework and hence model its SED in the stratified and the homogeneous medium. We conclude that although the  photons beyond the synchrotron limit can be interpreted by SSC process, the emission detected at hundreds of GeVs is due to the closeness  and the set of favorable parameter values of this burst. We conclude that low-redshift GRBs described under favourable set of parameters as found in GRB 190114C could be detected at hundreds of GeVs, and also afterglow transitions would allow that VHE emission  could be observed for longer periods.  The results of our afterglow model in the homogeneous medium is consistent with the results reported in \cite{2019arXiv190508285D}.  In our case, the SSC emission is below the KN regime $868.1\, {\rm GeV}$ with a bulk and electron Lorentz factors of $101.2$ and $1.9\times10^{-4}$, respectively.   It is worth noting that the parameters obtained with the MCMC code from the broadband modeling of the multi-wavelength observations may change somewhat when the KN effects are considered, but the SSC emission itself will not be strongly affected.
%
%
%

\acknowledgements
We thank Xiang-Yu Wang and B.B Zhang for useful discussions. N. F. acknowledge  financial  support  from UNAM-DGAPA-PAPIIT  through  Grant IA 102019.  RBD acknowledges support from the National Science Foundation under Grant 1816694.
%
%
%

\clearpage

%
%
%

\clearpage

\appendix
The proportionality constants in the stratified medium, eq. (\ref{ssc_wind}), are 
{\small
\bary\nonumber
F^{\rm W}_{\rm 1, f}&\simeq& 2.3\times 10^{-5}\,{\rm mJy} \,\left(\frac{1+Y_{\rm Th}}{3}\right)^{\frac43}\,\left(\frac{1+z}{1.5} \right)^3\,\xi^4 \varepsilon_{\rm B,-4}^\frac53 \,A^{-4}_{W,-1}\,D^{-2}_{27}\,E^{-\frac13}_{53.5}\,,\cr
F^{\rm W}_{\rm 2, f}&\simeq& 2.1\times 10^{-3}\,{\rm mJy} \,\left(\frac{1+Y_{\rm Th}}{3}\right)^{-2}  \,\left(\frac{1+z}{1.5} \right)^\frac12\,\xi^{-\frac14}\,  \varepsilon_{\rm B,-4}^{-\frac{5}{4}}\,D^{-2}_{27}\,A^\frac14_{W,-1}\,E^\frac12_{53.5}\,,\cr
F^{\rm W}_{\rm 3, f} &\simeq& 2.5\times 10^{-7}\,{\rm mJy}\,  g(p)^{2(p-1)}  \left(\frac{1+Y_{\rm Th}}{3}\right)^{-2} \left(\frac{1+z}{1.5} \right)^\frac{p}{2}\, \xi^{\frac{p-2}{4}}\,\varepsilon_{\rm B,-4}^\frac{p-6}{4}\,\varepsilon_{e,-2}^{2(p-1)}\, A^{\frac{2-p}{4}}_{W,-1} \,D^{-2}_{27}\,E^\frac{p}{2}_{53.5}\,, \cr
F^{\rm W}_{\rm 1,s}&\simeq& 4.2\times 10^{-2}\,{\rm mJy}\,  g(p)^{-\frac43} \,\left(\frac{1+z}{1.5} \right)^\frac53\,\xi^{-\frac83}\, \varepsilon_{\rm B,-4}^{\frac13}\,\varepsilon_{e,-2}^{-\frac{4}{3}}\,A^\frac83_{W,-1}\, D^{-2}_{27}\,E^{-\frac13}_{53.5}\,,\cr
F^{\rm W}_{\rm 2, s}&\simeq&8.6\times 10^{-10}\,{\rm mJy}\,  g(p)^{2(p-1)}\left(\frac{1+z}{1.5} \right)^\frac{p+3}{2}\, \xi^{\frac{p-11}{4}}\, \varepsilon_{\rm B,-4}^\frac{p+1}{4}\,\varepsilon_{e,-2}^{2(p-1)}\,A^{\frac{11-p}{4}}_{W,-1}\,D^{-2}_{27}\,E^\frac{p-1}{2}_{53.5}\,,\cr
\eary
}
with $F^{\rm W}_{\rm 3, s}=F^{\rm W}_{\rm 3, f}$.  The proportionality constants in the homogeneous medium, eq. (\ref{ssc_ism}), are 
{\small
\bary\nonumber
F^{\rm H}_{\rm 1, f}&\simeq& 2.3\times 10^{-5}\,{\rm mJy} \,\left(\frac{1+Y_{\rm Th}}{2.5}\right)^{\frac43}\,\, \varepsilon_{\rm B,-4}^\frac53 \,n^{-2}\,D^{-2}_{27}\,E^\frac53_{53.5}\,,\cr
F^{\rm H}_{\rm 2, f}&\simeq&  2.5\times 10^{-6}\,{\rm mJy} \,\left(\frac{1+Y_{\rm Th}}{2.5}\right)^{-2}  \,\left(\frac{1+z}{1.5} \right)^\frac38\, \varepsilon_{\rm B,-4}^{-\frac{5}{4}}\,D^{-2}_{27}\,n^\frac18\,E^\frac58_{53.5}\,,\cr
F^{\rm H}_{\rm 3, f} &\simeq& 2.7\times 10^{-10}\,{\rm mJy}\,  g(p)^{2(1-p)}  \left(\frac{1+Y_{\rm Th}}{2.5}\right)^{-2} \left(\frac{1+z}{1.5} \right)^\frac{5p-2}{8}\,\varepsilon_{\rm B,-4}^\frac{p-6}{8}\,\varepsilon_{e,-2}^{2p-2}\,D^{-2}_{27}\,E^\frac{3p+2}{8}_{53.5}\,, \cr
F^{\rm H}_{\rm 1,s}&\simeq& 4.6\times 10^{-3}\,{\rm mJy}\,  g(p)^{\frac43} \,\left(\frac{1+z}{1.5} \right)^\frac13\,\varepsilon_{\rm B,-4}^{\frac13}\,\varepsilon_{e,-2}^{-\frac{4}{3}}\,n^\frac43\, D^{-2}_{27}\,E_{53.5}\,,\cr
F^{\rm H}_{\rm 2, s}&\simeq&7.4\times 10^{-11}\,{\rm mJy}\,  g(p)^{2(p-1)}\left(\frac{1+z}{1.5} \right)^\frac{5p+1}{8}\,\varepsilon_{\rm B,-4}^\frac{p+1}{4}\,\varepsilon_{e,-2}^{2(p-1)}\,n^{\frac{11-p}{8}}\,D^{-2}_{27}\,E^\frac{3p+7}{8}_{53.5}\,,\cr
\eary
}
with $F^{\rm H}_{\rm 3, s}=F^{\rm H}_{\rm 3, f}$.

\end{document}